\documentclass[twocolumn,showpacs,showkeys,preprintnumbers,amsmath,amssymb,aps,floatfix]{revtex4}
\pdfoutput=1
\usepackage{color}
\usepackage{indentfirst}
\usepackage{graphicx}
      \def\di{\displaystyle}
      
      \def\bS{{\bf S}}
      \def\bl{{\bf l}}
      \def\bp{{\bf p}}
      
      \def\br{{\bf r}}

      \def\F{{\cal F}}
      \def\H{{\cal H}}

      \def\L{{\cal L}}
      
      \def\P{{\cal P}}
      \def\R{{\cal R}}

      \def\e{{\rm e}}

      \def\d{{{\downarrow\uparrow}}}
\begin{document}

\title{ Experimental status of the nuclear spin scissors mode}

\author{ E. B. Balbutsev\email{balbuts@theor.jinr.ru}, I. V. Molodtsova\email{molod@theor.jinr.ru}}
\affiliation{ Joint Institute for Nuclear Research, 141980 Dubna, Moscow Region,Russia }
\author{ P. Schuck}
\affiliation{Institut de Physique Nucl\'eaire, IN2P3-CNRS, Universit\'e Paris-Sud,
F-91406 Orsay C\'edex, France;\\
Laboratoire de Physique et Mod\'elisation des Milieux Condens\'es,
CNRS and Universit\'e Joseph Fourier,
25 avenue des Martyrs BP166, F-38042 Grenoble C\'edex 9, France}

\begin{abstract}
With the Wigner Function Moments (WFM) method the scissors mode of
the actinides and rare earth nuclei are investigated. The unexplained experimental fact that 
in $^{232}$Th a double
hump structure is found finds a natural explanation within WFM. It is
predicted that the lower peak corresponds to an isovector spin scissors mode
whereas the higher lying states corresponds to the conventional isovector orbital
scissors mode. The experimental situation is scrutinized in this respect concerning practically all results
of $M1$ excitations.

\end{abstract}


\maketitle

\section{Introduction}

\begin{figure}[h]
\centering\includegraphics[width=0.5\columnwidth]{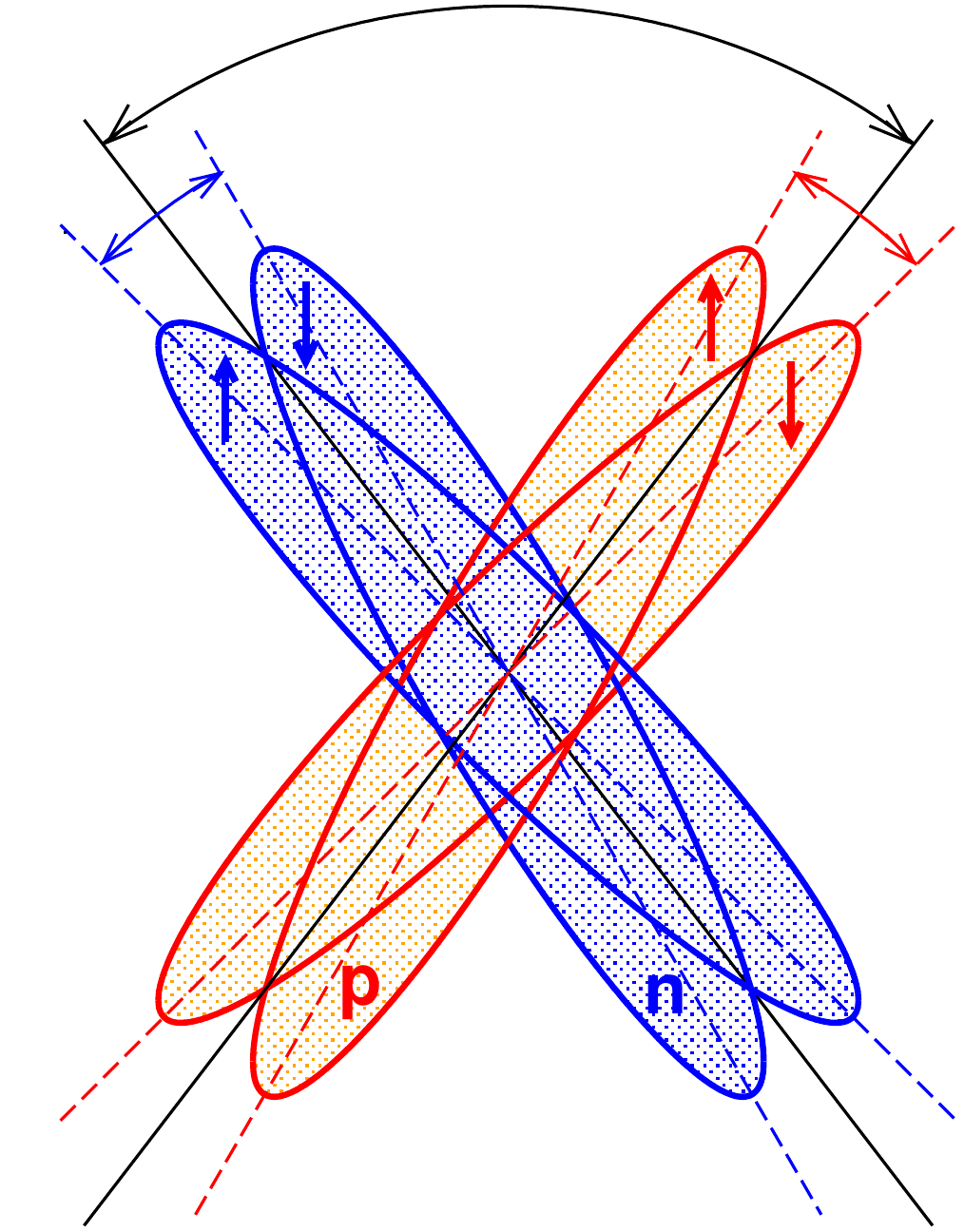}
\caption{(Color online) Pictorial representation of two intermingled scissors: the 
orbital scissors (neutrons versus protons) + spin scissors (spin-up 
nucleons versus spin-down nucleons). Arrows inside of ellipses show the 
direction of spin projections.
 p -- protons, n -- neutrons.}
\label{fig1}\end{figure}

 In a recent paper \cite{BaMo} the Wigner Function Moments (WFM) 
or phase space moments method  was applied for the first time 
to solve the TDHF equations including spin dynamics.
As a first step, only the spin orbit interaction was included in the
consideration, as the most important
one among all possible spin dependent interactions because it enters 
into the mean field. 
 The most remarkable result was the prediction of a new type
of nuclear collective motion: rotational oscillations of "spin-up"
nucleons with respect of "spin-down" nucleons (the spin scissors mode). 
It turns out that the experimentally 
observed group of peaks in the energy interval $2-4$ MeV corresponds 
very likely to
two different types of motion: the orbital scissors mode and this new kind 
of mode, i.e. the spin scissors mode. The pictorial view of these two intermingled scissors
is shown in Fig.~\ref{fig1}. It just shows the  generalization of the 
classical picture for 
the orbital scissors (see, for example, \cite{Lo2000,Heyd}) to include the spin scissors modes.
\begin{figure}[t!]
\centering\includegraphics[width=0.8\columnwidth]{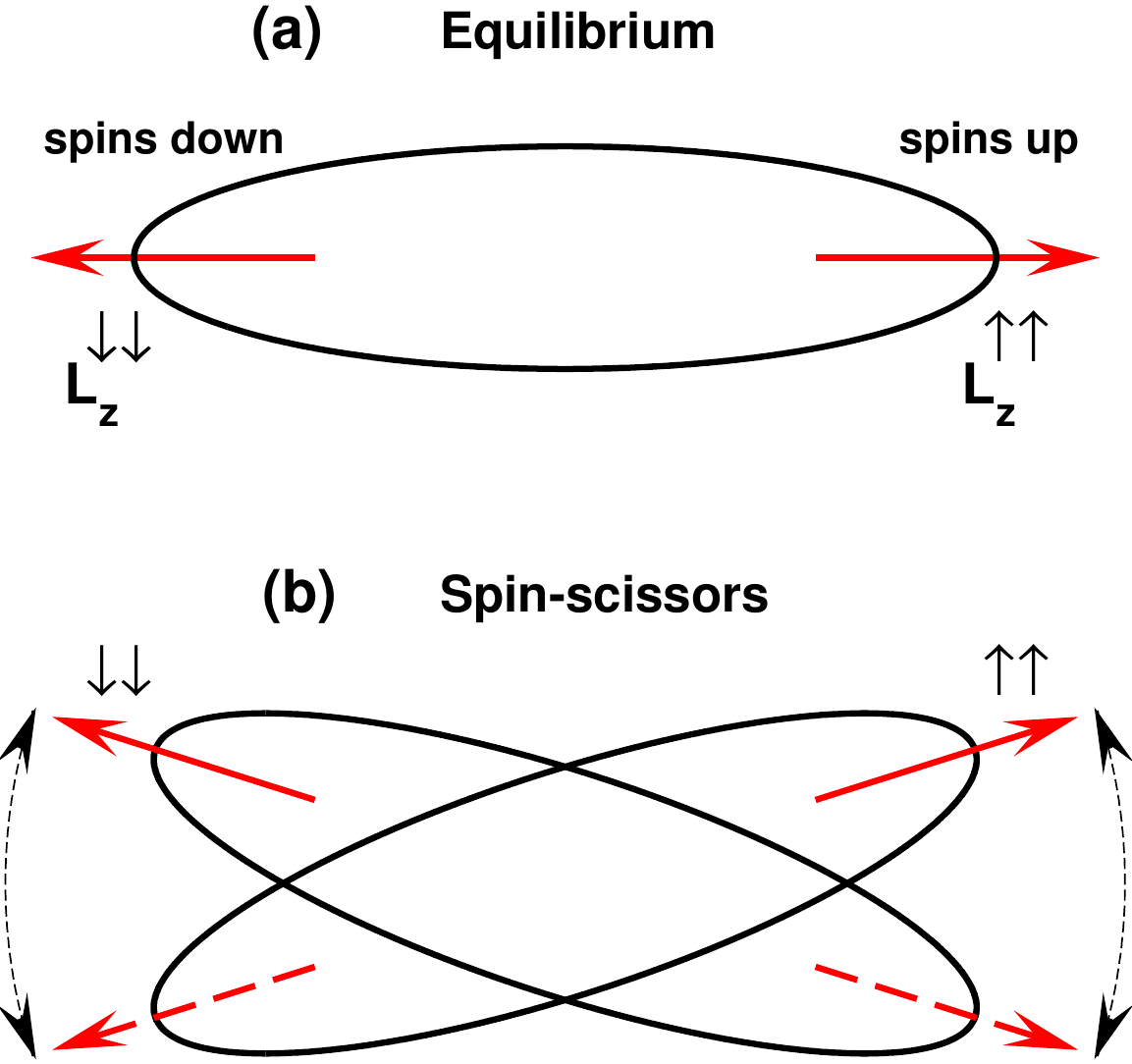}
\caption{(Color online) (a) Protons with spins $\uparrow$ (up) and $\downarrow$ (down) having nonzero orbital
angular momenta at equilibrium. (b) Protons from Fig.(a) vibrating against one-another.}
\label{figSch}\end{figure}
\begin{figure}[h!]
\centering\includegraphics[width=\columnwidth]{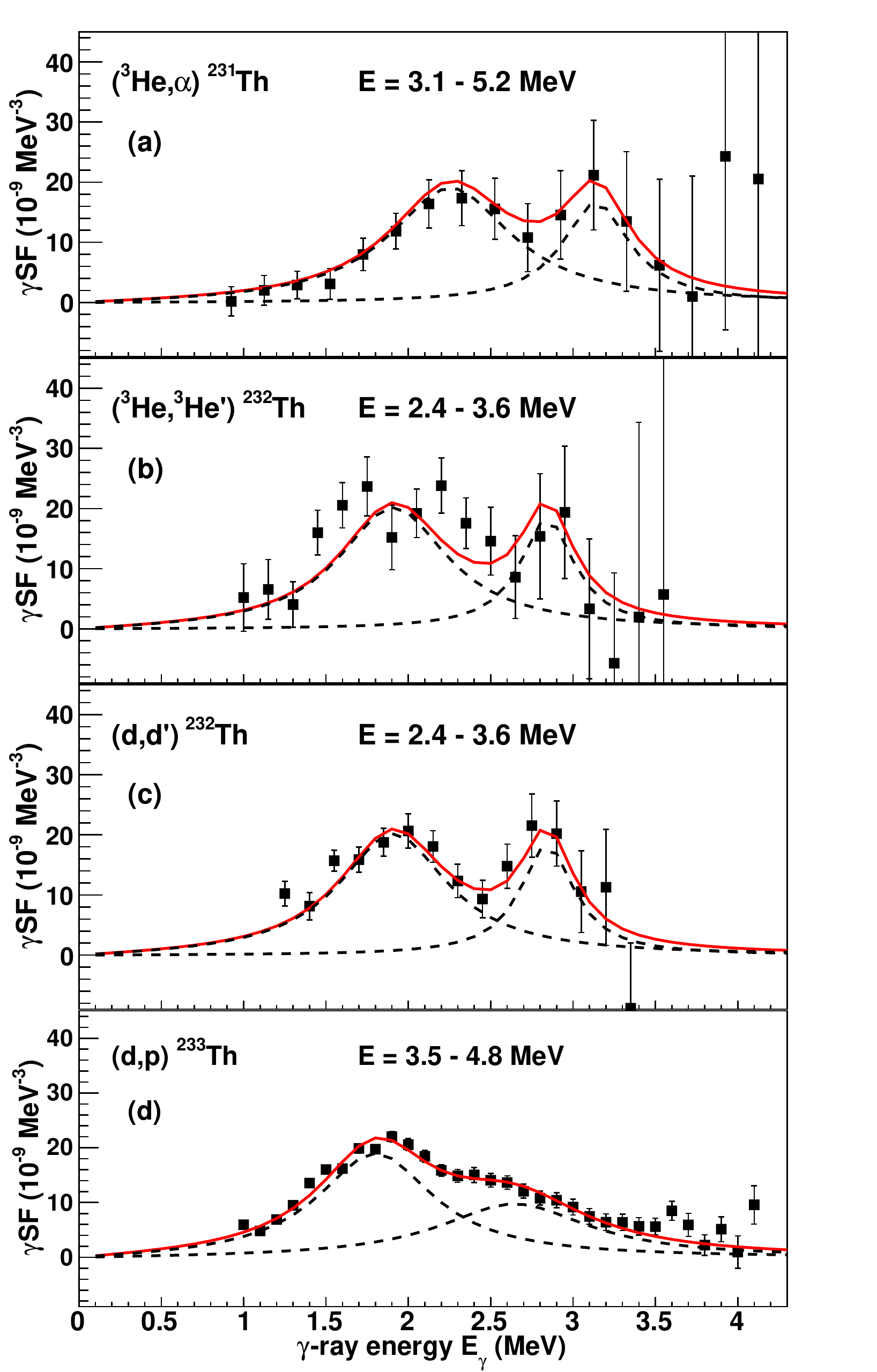}
\includegraphics[width=0.99\columnwidth]{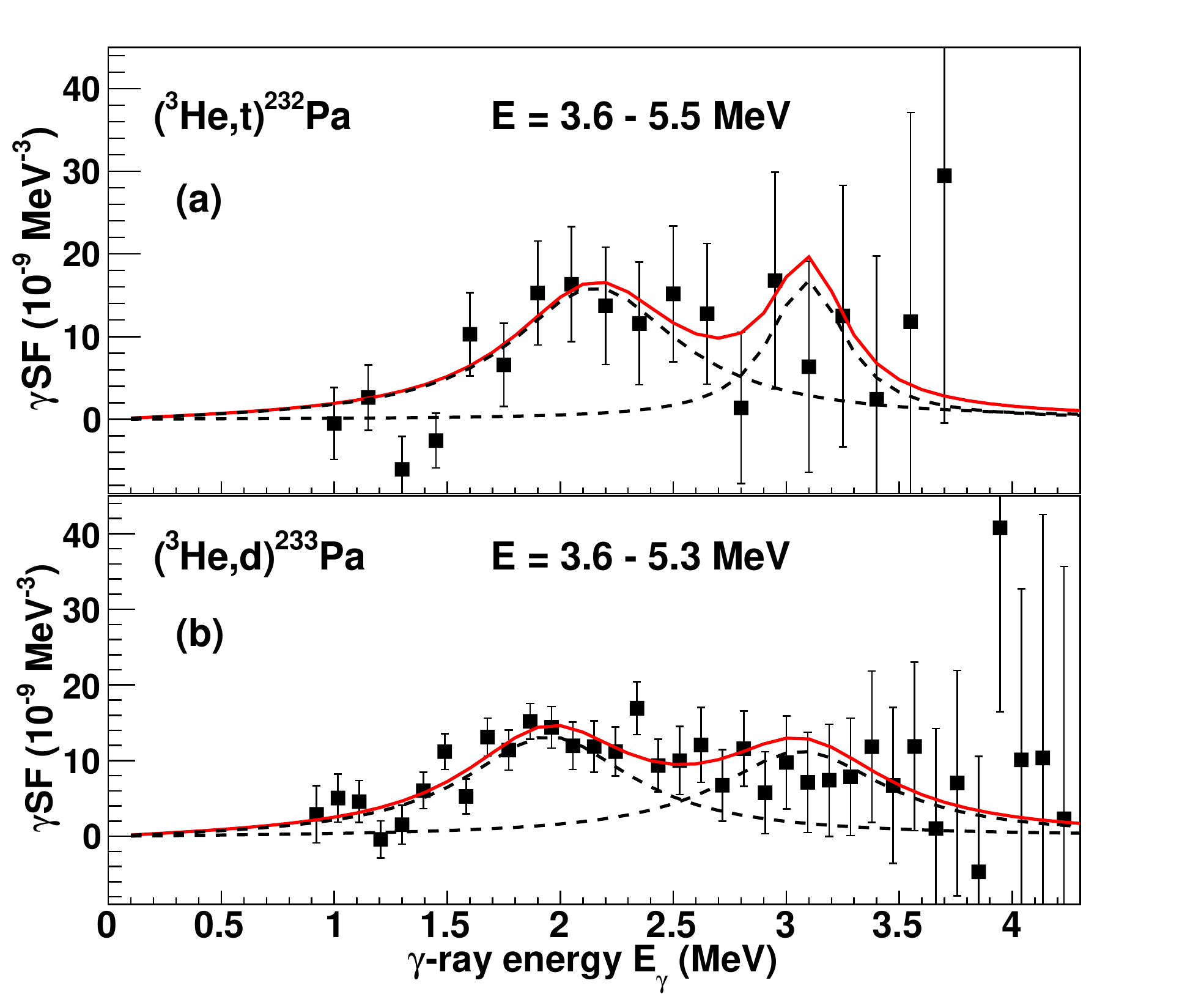}
\caption{(Color online) Radiative Strength Functions (RSF) for scissors resonance observed by
Oslo group \cite{Siem2} for $^{231-233}$Th and $^{232-233}$Pa.
This figure is taken from the paper \cite{Siem2}.}
\label{figSiem}\end{figure}
\begin{figure}[t!]
\centering
\includegraphics[width=\columnwidth]{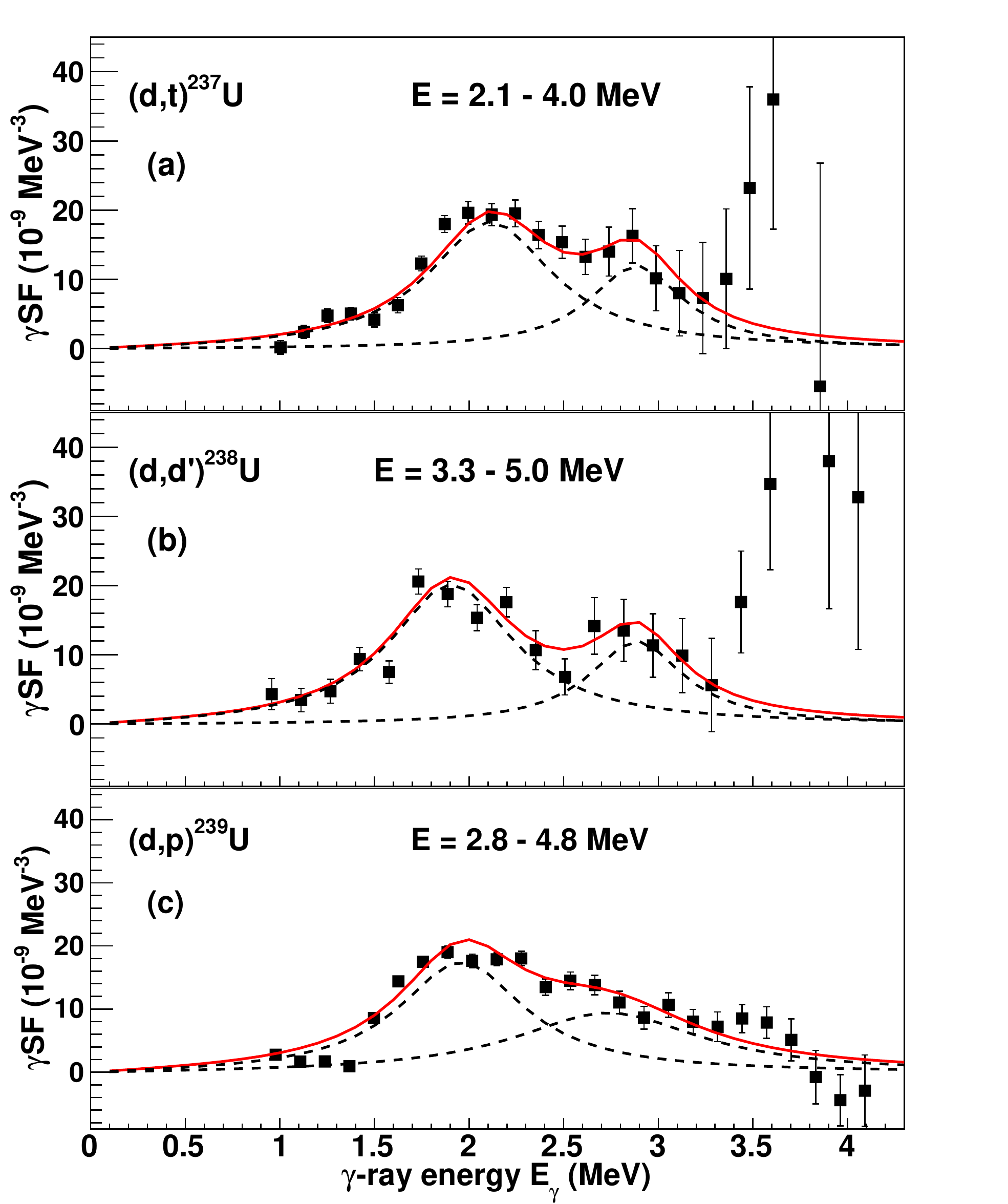}
\caption{(Color online) Same as Fig.~\ref{figSiem} for $^{237-239}$U.
This figure is taken from the paper \cite{Siem2}.
}
\label{U}\end{figure}

In Ref. \cite{BaMoPRC} the influence of the spin-spin interaction on the scissors modes was studied. 
It was found that such interaction does not push the predicted mode strongly up in energy. 
It turned out that the spin-spin interaction does not change the general
picture of the positions of excitations described in \cite{BaMo} pushing all levels
up proportionally to its strength without changing their order. The most interesting
result concerns the $B(M1)$ values of both scissors
-- the spin-spin interaction strongly redistributes $M1$ strength in  favour
of the spin scissors mode without changing their summed strength.

A generalization of the WFM method which takes into account spin degrees
of freedom and pair correlations simultaneously was outlined in 
\cite{BaMoPRC2}, where the rare earth nuclei were considered. As a result the agreement between theory and
experiment in the description of nuclear scissors modes was improved
considerably.
The decisive role in the substantial improvement of results is played by the 
anti-aligned spins~\cite{BaMoPRC2}. 
It was shown that the ground state nucleus consists 
of two equal parts having nonzero angular momenta with opposite directions, which compensate each other resulting in the zero total angular momentum. 
This is graphically depicted in Fig.~\ref{figSch}(a). 
On the other hand, when the opposite angular momenta become tilted, one excites the system and the opposite 
angular momenta are vibrating with a tilting angle, see Fig.~\ref{figSch}(b). 
It is rather obvious from Fig.~\ref{figSch} that these tilted vibrations 
happen separately in each of the neutron and proton lobes. 
These spin-up against spin-down motions certainly influence the 
excitation of the spin scissors mode. 
 
The aim of the present paper is to find the  experimental confirmation
of our prediction: the splitting of low lying ($E<4$ MeV) $M1$ excitations
in two groups corresponding to spin and orbital
scissors, the spin scissors being lower in energy and stronger in 
transition probability $B(M1)$. 
It turns out that a similar phenomenon was observed and
discussed by experimentalists already at the beginning of the "scissors era". For example, we cite from the paper of 
C.~Wesselborg {\it et al.}~\cite{Wessel}: "The existence of the two groups poses the question 
whether they arise from one mode, namely from scissors mode, or
whether we see evidence for two independent collective modes."
We already touched this problem in the
paper~\cite{BaMoPRC}, where we have found, that our theory explains
quite naturally the experimental results of Oslo group~\cite{Siem}
for~$^{232}$Th. 

We, therefore, will concentrate to a large extent on the explanation of
the experiments in the actinides, see Sec.~\ref{s2}. In Sec.~\ref{s3}, we will try
to see whether in the rare earth nuclei there are also signs of a double
hump structure.  
In Sec.~\ref{s7} we will give our conclusions.
The description of the WFM method and mathematical details are given in Appendix~\ref{AppA}.

\section{The situation in the actinides}\label{s2}

Guttormsen {\it et al} \cite{Siem} have studied 
deuteron and $^3$He-induced reactions on $^{232}$Th and found in the 
residual nuclei $^{231,232,233}$Th and $^{232,233}$Pa
"an unexpectedly strong integrated strength of $B(M1)=11-15~\mu_N^2$ in the 
$E_\gamma=1.0-3.5$~MeV region". The $B(M1)$ force in most nuclei shows 
evident splitting into two Lorentzians. "Typically, the experimental
splitting is $\Delta\omega_{M1}\sim 0.7$~MeV, and the ratio of the 
strengths between the lower and upper resonance components is
$B_L/B_U\sim 2$". 
Seeing this obvious splitting the question is raised: "What is
the nature of the splitting?" Their attempt to explain the 
splitting by a $\gamma$-deformation has failed. 
\begin{figure}[h]
\centering\includegraphics[width=\columnwidth]{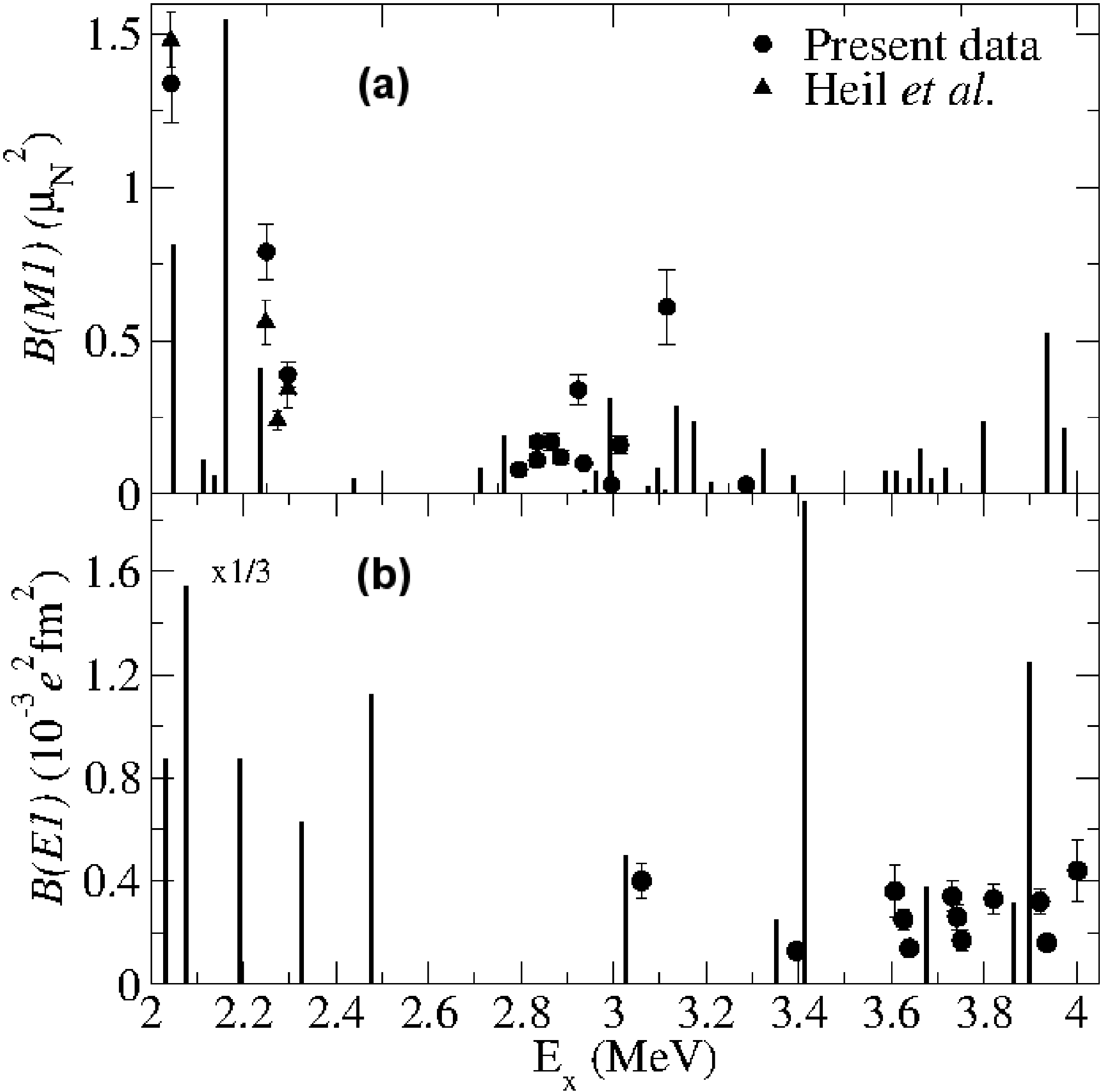}
\caption{Comparison of experimental and calculated $B(M1)$ (panel (a)) in 
$^{232}$Th. Heil {\it et al} \cite{Heil}. The solid line represents QRPA calculations \cite{Kuliev}.
This figure is taken from the paper \cite{Adekola}; only panel (a) is necessary for us.}
\label{figAdekola}\end{figure}
To describe the observed value of
$\Delta\omega_{M1}$ the deformation $\gamma\sim 15^\circ$ is required, 
that leads to the ratio $B_L/B_U\sim 0.7$ in an obvious contradiction 
with experiment. The authors conclude that "the splitting may be 
due to other mechanisms". 
Later~\cite{Siem2} they reanalyzed their data for Th and Pa
with the result shown in Fig.~\ref{figSiem} and presented the results of new experiments for $^{237-239}$U
(Fig.~\ref{U})
with the conclusion: "The SR (Scissors Resonance)
displays a double-hump structure that is theoretically not understood."
\begin{table*}
\caption{\label{tab0} Scissors modes energies $E$ and
transition probabilities $B(M1)$. WFM -- the results of our calculations
for $^{232}$Th are compared with experimental
data: Exp.$^c$ -- Ref.~\cite{Adekola}, Exp.$^h$ -- Ref.~\cite{Siem2}.}
\begin{ruledtabular}
\begin{tabular}{lccclccc}
 $^{232}$Th  & \multicolumn{3}{c}{ $E$ (MeV)} & &\multicolumn{3}{c}{ $B(M1)\ (\mu_N^2)$}    \\
\cline{2-4} \cline{6-8}
    & WFM & Exp.$^c$ & Exp.$^h$ &  & WFM & Exp.$^c$ & Exp.$^h$ \\
\hline
 spin  sc.          & 2.30 & 2.15 & 1.95  & spin  sc.     & 2.40 & 2.52(26) & 6.5 \\
 orb.  sc.          & 2.93 & 2.99 & 2.85  & orb.  sc.     & 1.42 & 1.74(38) & 3.0 \\ \hline\hline
 $\Delta E$         & 0.63 & 0.84 & 0.8   & $B_L/B_U$     & 1.69 & 1.45 & 2.2 \\ \hline  
 $E_{\rm cent}$     & 2.54 & 2.49 & 2.2   & $B_\Sigma$    & 3.82 & 4.26(64) & 9.5 \\ 
\end{tabular}
\end{ruledtabular}
\end{table*}

When the paper \cite{BaMoPRC} with our explanation of the two humps nature of
SR was published, P. von Neumann-Cosel attracted our attention to the paper by
A. S. Adekola {\it et al} \cite{Adekola} who have studied the scissors mode in
the ($\gamma,\gamma'$) reaction on the same nucleus $^{232}$Th one year earlier.
It turns out that these authors have also obtained the scissors mode splitting (see
Fig.~\ref{figAdekola}), but did not pay further attention to it.
\begin{figure}[b!]
\centering\includegraphics[width=\columnwidth]{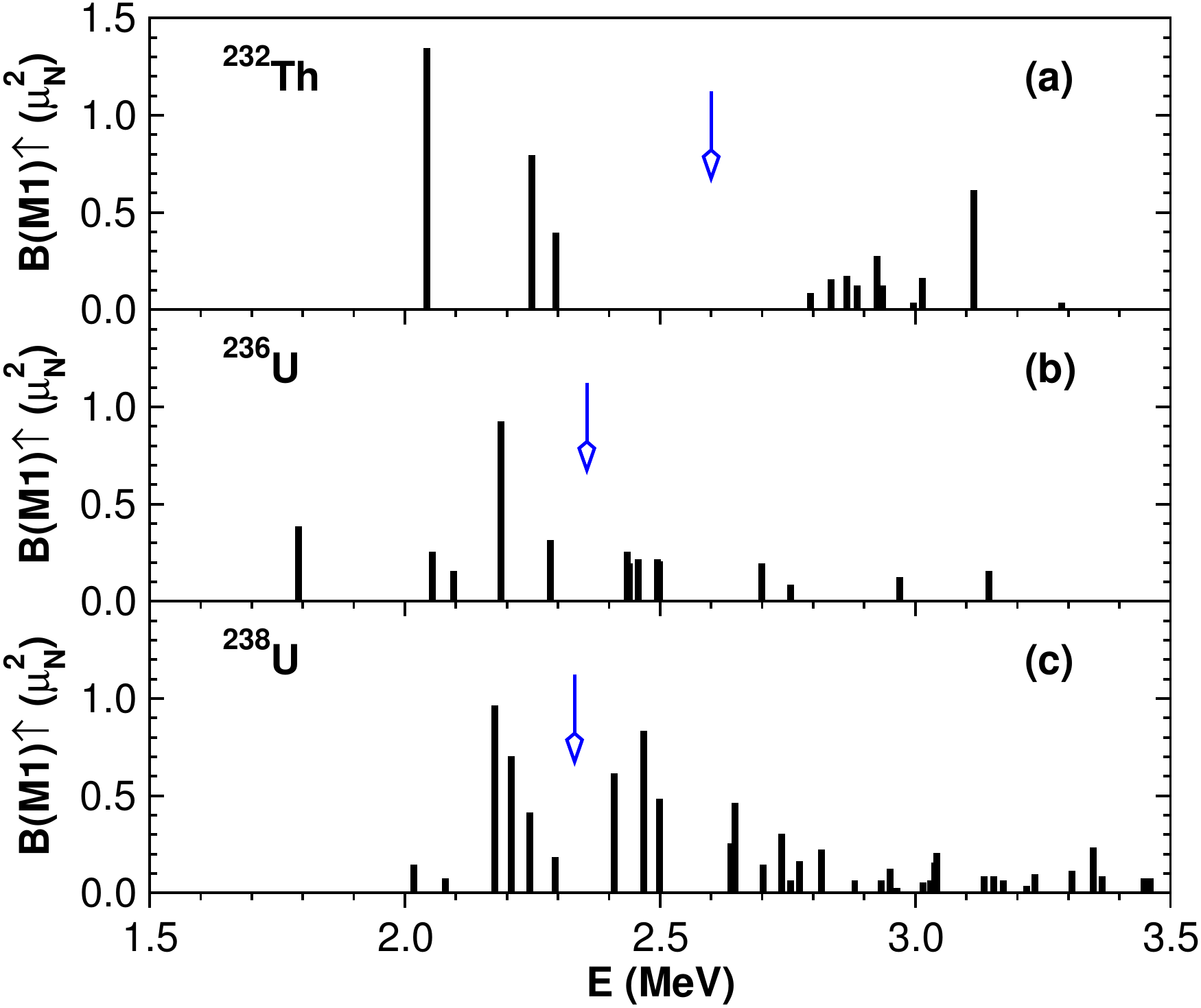}
\caption{(Color online) The experimentally observed spectra of $1^+$ excitations: (a)$^{232}$Th -- \cite{Adekola},
 (b)$^{236}$U -- \cite{U236},  (c)$^{238}$U -- \cite{Hammo}. 
The point of the division of the spectrum in two parts is shown by a blue arrow.
}\label{fig5}
\end{figure}
\begin{table*}
\caption{\label{tab_actinids}
Energy centroids $E$ and summed transition probabilities $B(M1)$ of the spin and orbital scissors. 
Experimental data: $^{232}$Th -- \cite{Adekola}, $^{236}$U -- \cite{U236}, $^{238}$U -- \cite{Hammo}.
The information for $^{236}$U contains second line, which is obtained by taking into account $1^{+}$ 
states together with $1^{\pi}$ states with $\pi$ unknown. }
\begin{ruledtabular}
\begin{tabular}{lcccccccccclc}
 Nuclei & 
\multicolumn{6}{c}{ $E$ (MeV)}    &
\multicolumn{6}{c}{ $B(M1)\ (\mu_N^2)$}    \\
\cline{2-7} \cline{8-13}
 &\multicolumn{2}{c}{spin scissors} &\multicolumn{2}{c}{orbital scissors}  &\multicolumn{2}{c}{centroid} & 
  \multicolumn{2}{c}{spin scissors} &\multicolumn{2}{c}{orbital scissors}  &\multicolumn{2}{c}{ $\sum$ } \\
\cline{2-3} \cline{4-5} \cline{6-7} \cline{8-9} \cline{10-11} \cline{12-13}  
 & Exp. & WFM & Exp. & WFM & Exp. & WFM & 
   Exp. & WFM & Exp. & WFM &\multicolumn{1}{c}{Exp.} & WFM\\
\hline   
 $^{232}$Th  & 2.15 & 2.30 & 2.99 & 2.93 & 2.49(37) & 2.54 & 2.52(26) & 2.40 & 1.74(38) & 1.42 & 4.26(64)& 3.82    \\  
 $^{236}$U   & 2.10 & 2.33 & 2.61 & 2.96 & 2.33 & 2.57 & 2.01(25) & 2.83 & 1.60(26) & 1.73 & 3.61(51)    & 4.56    \\
             & 2.12 &      & 2.63 &      & 2.35 &      & 2.26(29) &      & 1.80(31) &      & 4.06(60)    &         \\     
 $^{238}$U   & 2.19 & 2.36 & 2.76 & 3.00 & 2.58 & 2.61 & 2.46(31) & 3.18 & 5.13(89) & 2.02 & 7.59(1.20)  & 5.20    \\  
\end{tabular}                                                                                                         
\end{ruledtabular}
\end{table*}
The energy and $B(M1)$ values of the two humps of SR are shown in Table~\ref{tab0}.
As it is
seen, two experiments \cite{Siem2,Adekola}
demonstrate very good agreement in the description of
the splitting: $\Delta E$ and $B_L/B_U$. The big difference in $B(M1)$ values is
explained by the fact that in Oslo method one extracts the scissors mode from 
strongly excited (heated) nuclei, whereas in ($\gamma,\gamma'$) reactions one
deals with SR in cold (ground state) nuclei. WFM method describes small 
amplitude deviations from the ground state, so our results must be closer to
that of ($\gamma,\gamma'$) experiment that is confirmed in Table~\ref{tab0}. 
It is easy to see that the calculated energies $E$ and $B(M1)$ values of the spin and orbital scissors
are in very good agreement with the experimental \cite{Adekola} energy centroids 
and summarized $B(M1)$ values of the lower and higher groups of levels respectively.
Figure~\ref{figAdekola} contains also the results of QRPA calculations
of Kuliev {\it et al}~\cite{Kuliev} which are also very intreaguing. As in the 
experimental works they find a bunch of low-lying states in the region $2.0-2.3$ MeV 
and a second one in the region $2.7-3.4$ MeV. They also find a third bunch close to 4 MeV. 
Those results are very similar to ours in what concerns the two low
lying structures. It is very tempting to identify their low lying structure
with the spin scissors and the second structure with the orbital one.
However, the authors did not investigate their structures in those terms.
Indeed in an QRPA calculation it is not  evident to analyze what is
spin and what orbital scissors mode. 


It is necessary to stress that the solution of the set of dynamical equations
(\ref{iv}) gives only two low lying eigenvalues, which are interpreted as the centroid
energies of the spin scissors and the orbital scissors according to collective variables responsible for the generation of these eigenvalues (see, for example, 
Table I of our paper \cite{BaMoPRC}). That is why we can compare the results
of our calculations (two eigenvalues) only with the equivalent centroids of the 
experimental scissors spectra.

\begin{figure}[t!]
\centering\includegraphics[width=\columnwidth]{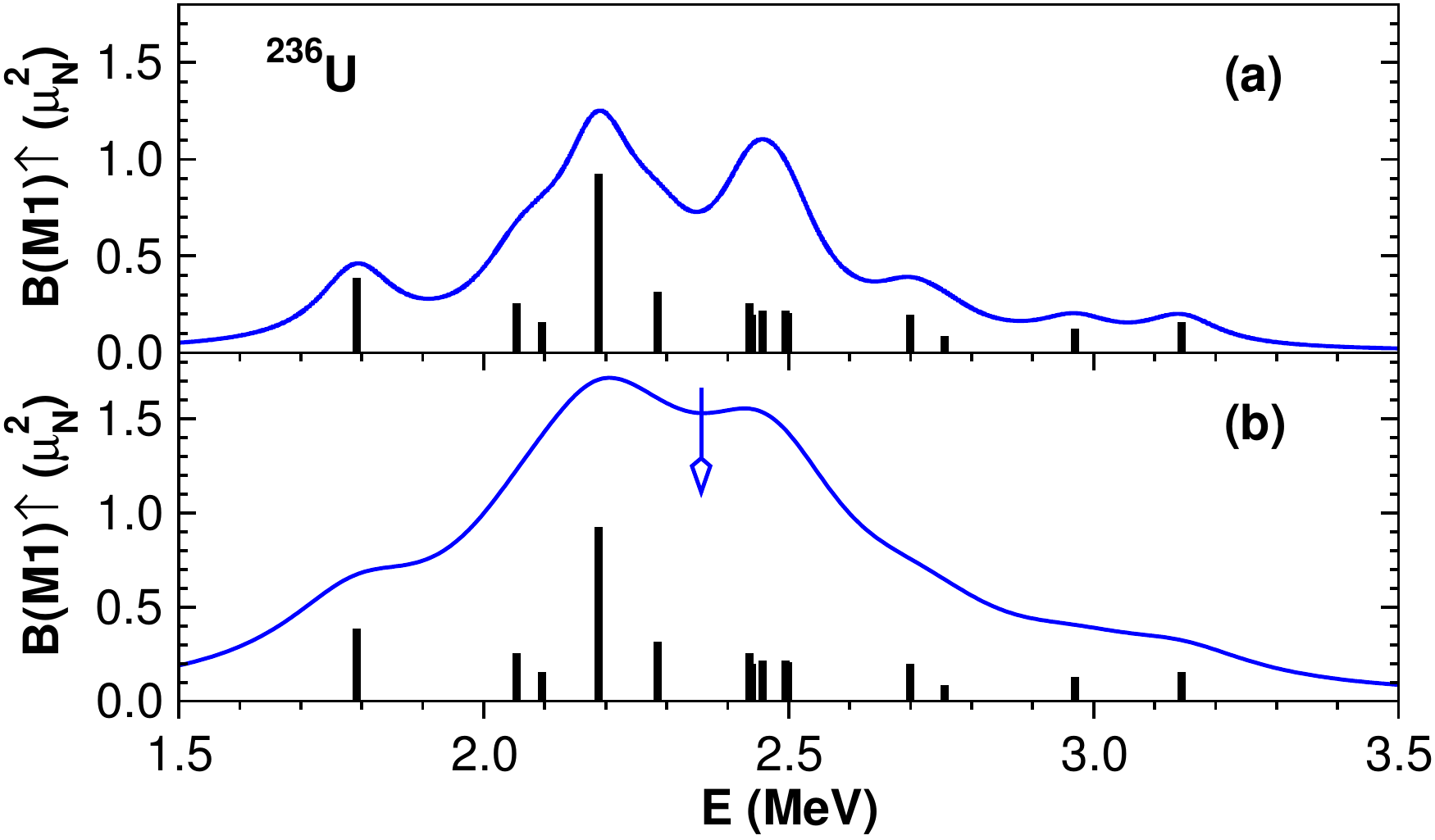}
\caption{(Color online) The experimentally observed spectrum of $1^+$ excitations of $^{236}$U -- \cite{U236} is divided in two parts with the help
of Lorentzian. The point of the division (the point of minimum between two
humps) is shown by a blue arrow.
The artificial widths of the spectrum lines used for folding in the case (a)
are smaller than the artificial widths used in the case (b).
}\label{fig6_0}
\end{figure}

The experimental spectra of $1^+$ excitations obtained in $(\gamma,\gamma')$ reactions for 
three actinide nuclei \cite{Adekola,U236,Hammo} are shown on Fig.~\ref{fig5}. 
The spectrum of $^{232}$Th can be divided in two groups with certainty. 
The division of spectra in two uranium nuclei is not so obvious. 
For example, there are four variants to divide the spectrum of $^{236}$U in two
groups: 1) to put the border (between two groups) into the energy interval 
$1.80 \leq E \leq 2.04$ MeV, 2) or into the interval
$2.30 \leq E \leq 2.41$ MeV, 3) or $2.51 \leq E \leq 2.69$ MeV,
4) or $2.78 \leq E \leq 2.95$ MeV.
Which interval to choose? To exclude any arbitrariness in the choice of the
proper interval, we apply a simple technical device. We folded the 
experimental spectra with a Lorentzian of increasing width. The width was increased until only two humps remained (see Fig.~\ref{fig6_0}). 
The blue arrows indicate the position of the minimum between the two humps. We want to stress the fact that the arrows are close to the position 
where the minima at finite temperature occur for $^{232}$Th and $^{238}$U. This gives some credit to our method to divide the spectra even at zero temperature 
into two humps, since it can be expected that if there is a two hump structure at finite temperature, there should be a similar one also at zero temperature.

\begin{figure}[b!]
\centering\includegraphics[width=\columnwidth]{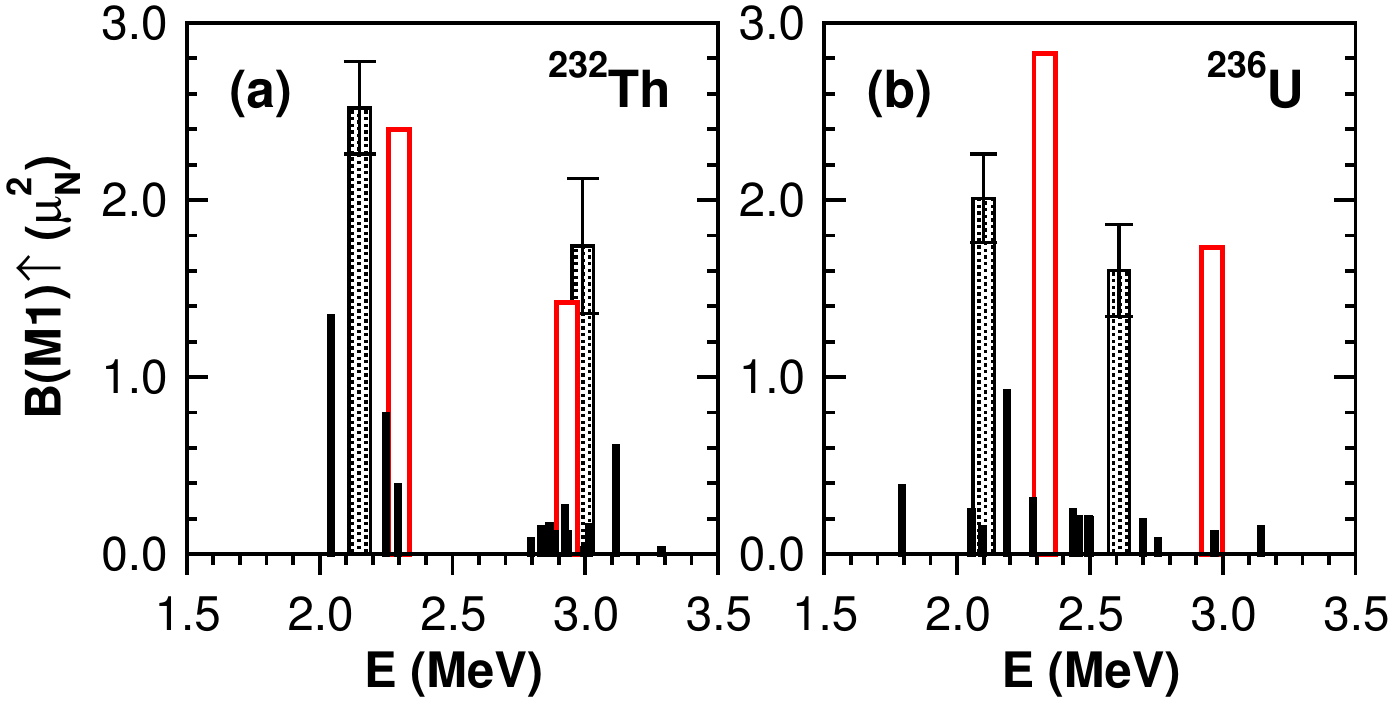}
\caption{(Color online) 
The centroids of experimentally observed spectra of $1^+$ excitations in 
$^{232}$Th (a) and $^{236}$U (b) (black rectangles with error bars) are compared with 
the results of calculations (red rectangles) 
of the spin and orbital scissors modes. In $^{236}$U, the theoretical results are slightly displaced to higher energy with respect to experiment.
}\label{WFM_00}
\end{figure}

Fig.~\ref{fig6_0} demonstrates the results of folding of $^{236}$U spectrum
with smaller (a) and bigger (b) values of the width of the Lorentzian. It is seen that the
spectrum of $^{236}$U can eventually be 
represented by a two-humps curve. 
Applying this procedure to $^{232}$Th and $^{238}$U we get the
results presented in Table~\ref{tab_actinids}. The points of the spectra
division are shown on Fig.~\ref{fig5} by blue arrows. The Table~\ref{tab_actinids} demonstrates rather good agreement between the theory and experiment
for both scissors in $^{232}$Th; the agreement in $^{236}$U 
can be characterized as acceptable. The situation is graphically displayed in Fig.~\ref{WFM_00}. One observes an unexpectedly large value of 
the summed $B(M1)$ for $^{238}$U in  comparison with that of $^{236}$U and 
$^{232}$Th and with the theoretical result. 
The possible reason of this discrepancy was indicated by the authors of
\cite{Hammo}: "$M1$ excitations are observed at approximately $2.0$ MeV $<E_{\gamma}< 3.5$ MeV with a strong concentration of $M1$ states around 2.5 MeV. ... 
The observed $M1$ strength may include states from both the scissors 
mode and the spin-flip mode, which are indistinguishable from each other based
exclusively on the use of the NRF technique." 
The most reasonable (and quite natural) place for the boundary between the scissors mode and the 
spin-flip resonance is located in the spectrum gap between 2.5 MeV and 2.62 MeV. The summed $M1$ strength 
of scissors in this case becomes $B(M1)=4.38\pm 0.5\ \mu_N^2$ in rather good agreement 
with  $^{236}$U and $^{232}$Th. This value is also not so far from the theoretical result. 
After dividing in spin and orbital scissors it gives $B(M1)_{\rm or}=1.92\ \mu_N^2$ in good agreement 
with the calculated value.

We want to stress again that the folding of spectra with Lorentzians is an artifact to divide a given spectrum into 
a lower lying and a higher lying group. The ensuing width of the two humps should not be interpreted as a true width. 
Also the specific choice of a Lorentzian has no significance. We could have chosen as well a Gaussian. 
Below we will apply exactly the same method for the case of Rare Earth nuclei to divide the spectra into two parts. 
As we will see, the case of Rare Earth nuclei is much harder but we will try and see. 
In conclusion, our method of separating the often pretty complex splitting patterns of the $M1$ strength into just two humps is an asumption 
which derives from our present theoretical description where besides the orbital scissors part, a spin scissors part appears.

It should be noted also that both scissors modes have an underlying orbital  nature, because 
both are generated by the same type of collective variables -- by the orbital 
angular momenta (the variables $\L_{\lambda\mu}$ in (\ref{Varis}) with $\lambda=\mu=1$).
 All the difference is that the "orbital" (conventional) scissors are generated 
by the counter-oscillations of the orbital angular momentum of protons with 
respect of the orbital angular momentum of neutrons, whereas  the "spin"  scissors  are generated by the counter-oscillations of the orbital angular  
momentum of nucleons having the 
spin projection "up" with respect of the orbital angular momentum  of nucleons 
having the spin projection "down". 
At the same time,  both scissors are strongly sensitive to the 
influence of the spin part of the magnetic dipole operator (\ref{Magnet_1})
\begin{equation}
\label{Magnet1}
\hat O_{11}=
\frac{\mu_N}{\hbar}\sqrt{\frac{3}{4\pi}}\left[g_s\hat S_{1}
+g_l\hat l_1\right].
\end{equation}
The Table~\ref{tab1} demonstrates the sensitivity of both scissors to the spin-dependent part of nuclear forces: 
the moderate constructive interference of the orbital and  spin contributions 
in the case of the spin scissors mode and their very strong 
destructive interference in the case of the orbital scissors mode.
\begin{table}[t!]
\caption{\label{tab1} Scissors modes energies $E$ and
transition probabilities $B(M1)$ for $^{232}$Th. The results of calculations with and without the spin
part of a dipole magnetic operator (\ref{Magnet1})
($g_s \neq 0$ and $g_s=0$ respectively).}
\begin{ruledtabular}
\begin{tabular}{ccccccc}
             & \multicolumn{3}{c}{ $E$ (MeV)} & \multicolumn{3}{c}{ $B(M1)$ ($\mu_N^2$) } \\
\cline{2-4}\cline{5-7}    
 & spin   & orb.    &   centroid  & spin & orb.  & $\sum$   \\ 
\hline 
 $g_s\neq 0$  & 2.30 & 2.93 & 2.54 & 2.40 & 1.42 & 3.82 \\
 $g_s = 0$    & 2.30 & 2.93 & 2.87 & 0.74 & 7.69 & 8.43 \\ 
\end{tabular}
\end{ruledtabular}
\end{table}

\noindent  
\begin{figure*}
\includegraphics[width=0.333\textwidth]{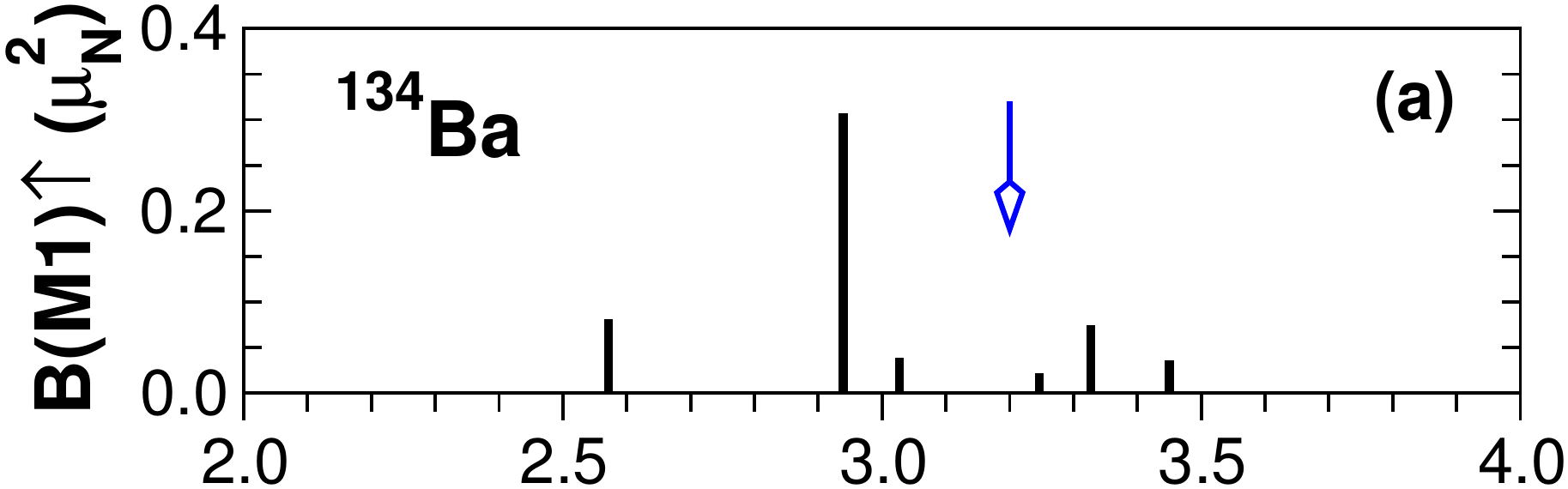}\\
\includegraphics[width=0.333\textwidth]{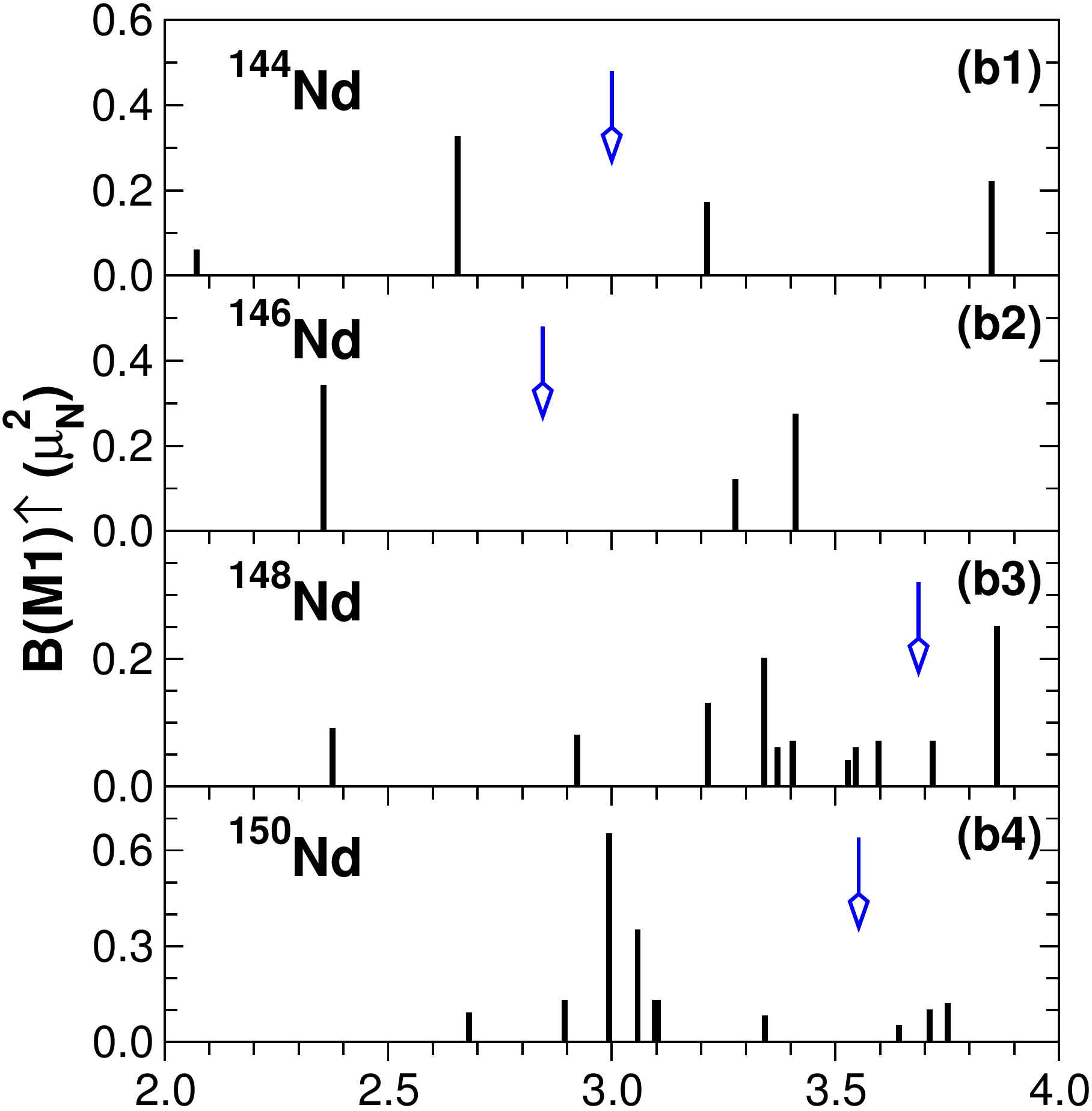}\hspace*{1mm}\includegraphics[width=0.305\textwidth]{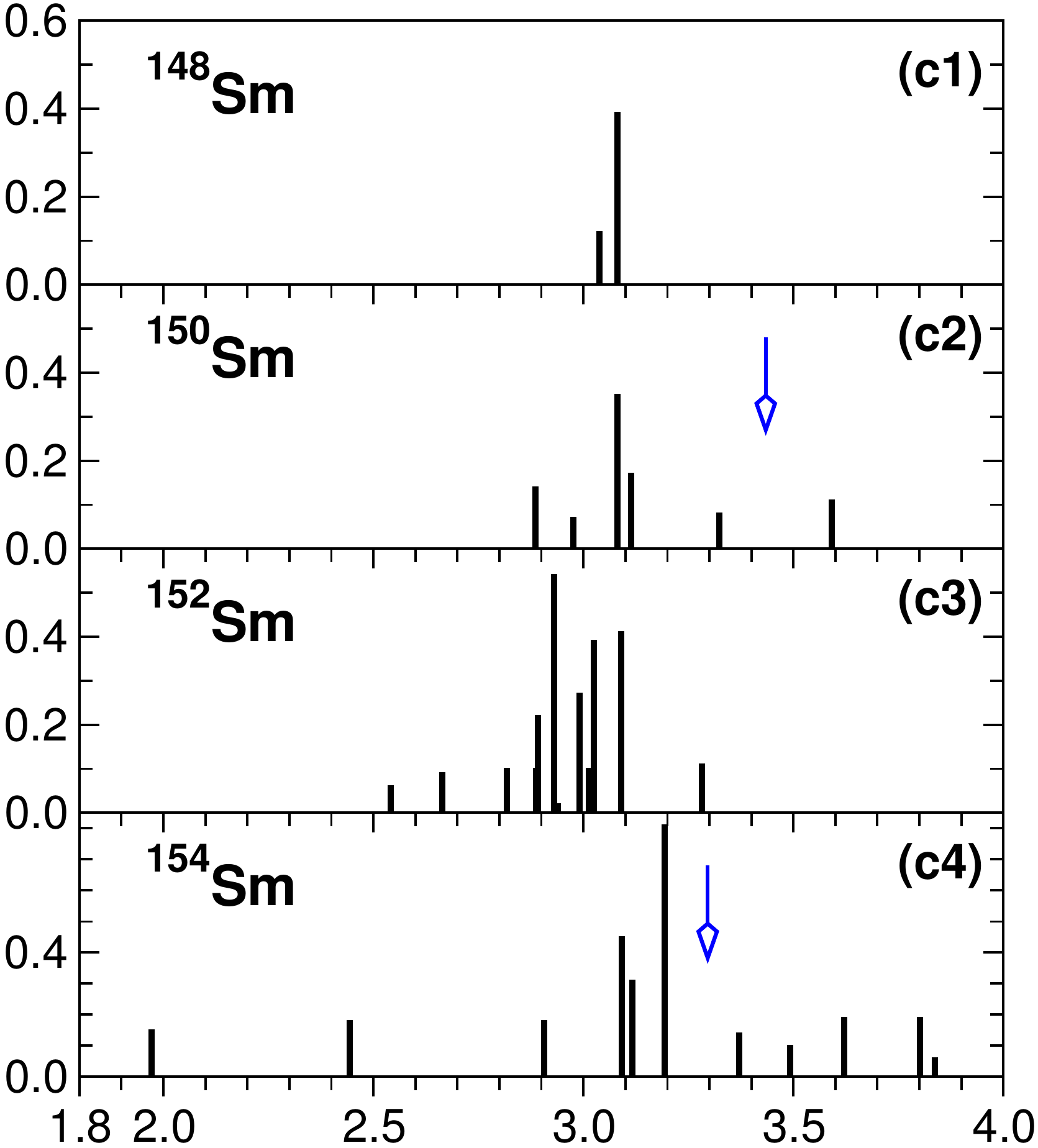}\hspace*{1mm}\includegraphics[width=0.305\textwidth]{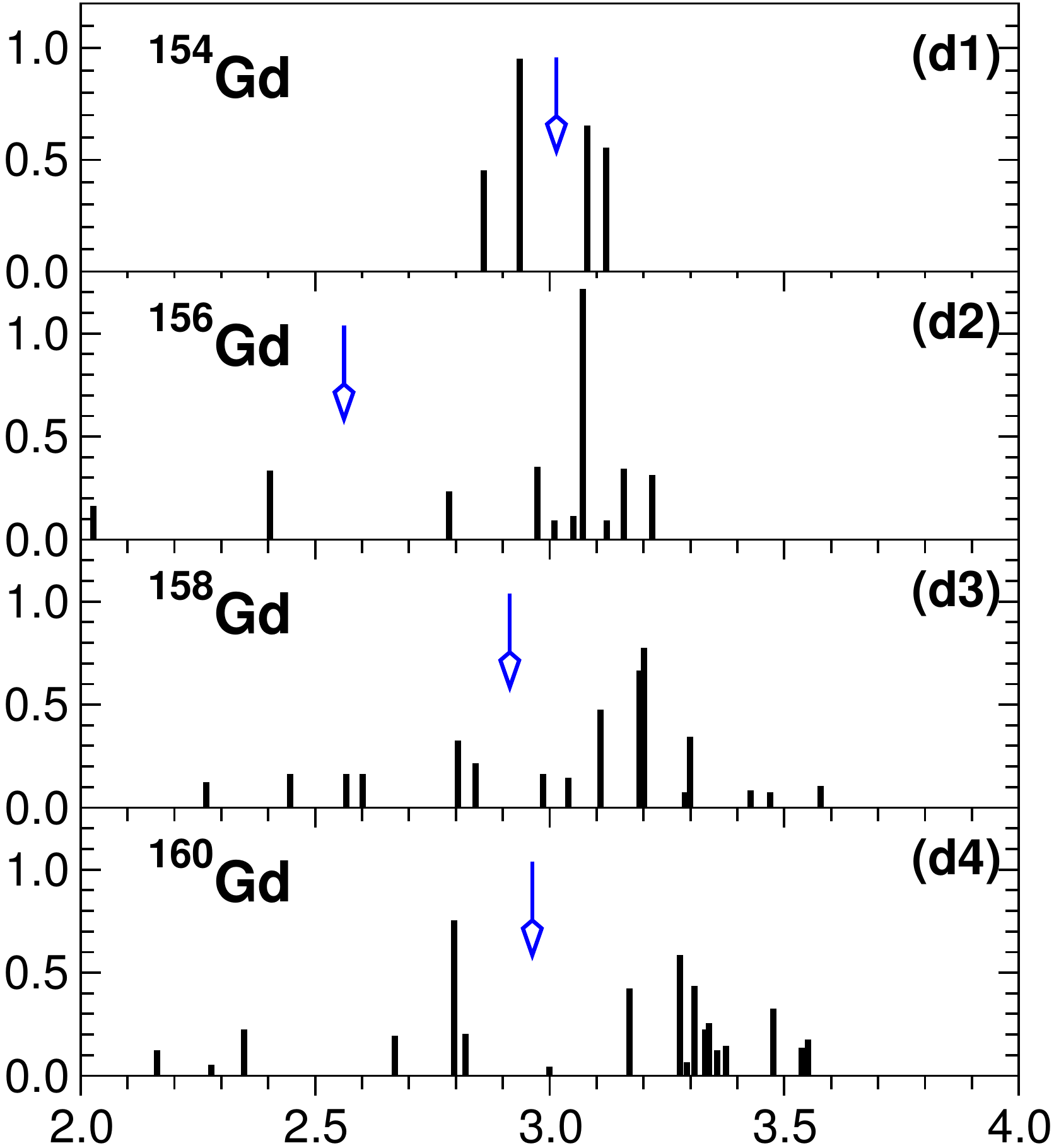}
\includegraphics[width=0.333\textwidth]{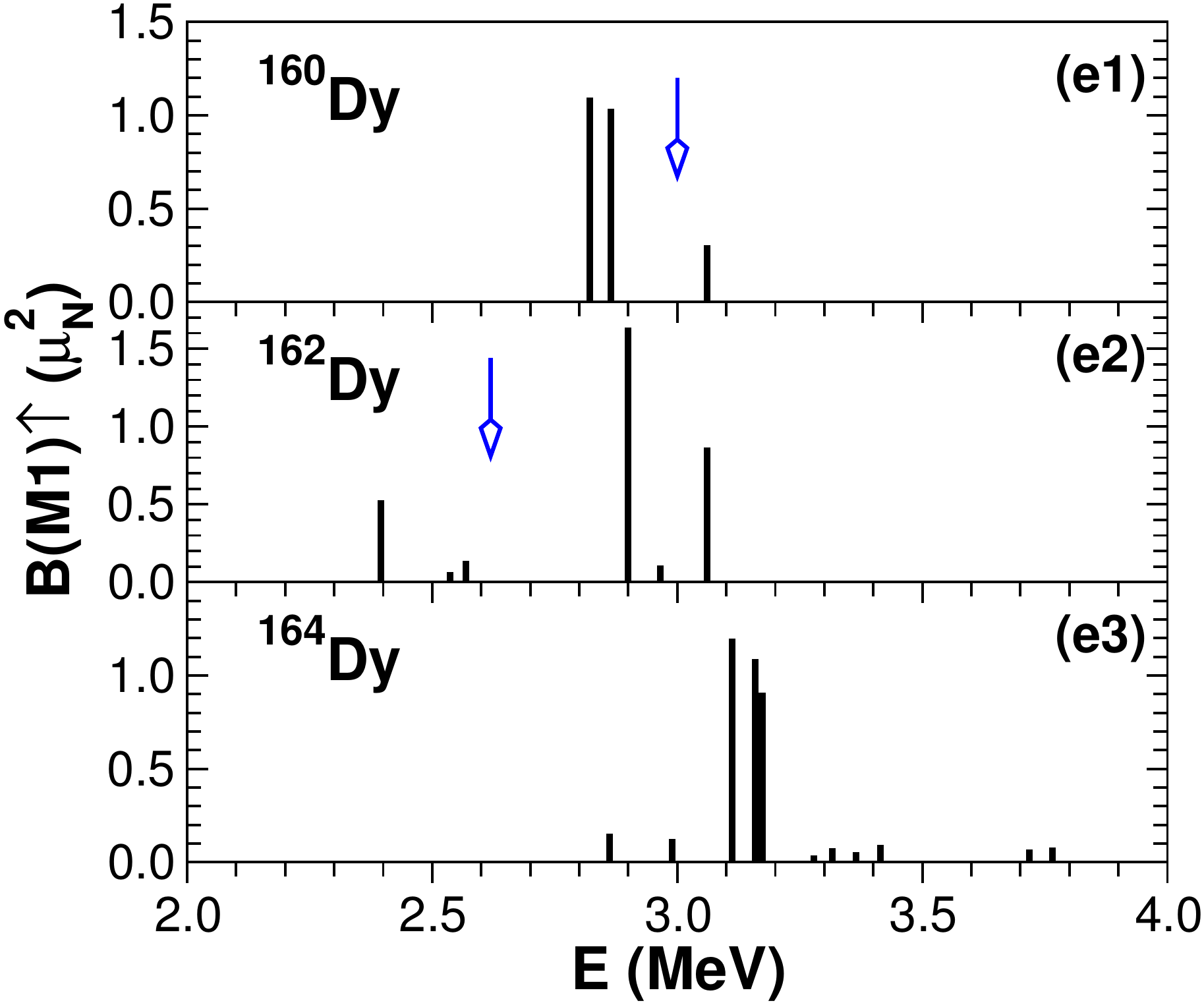}\hspace*{1mm}\includegraphics[width=0.305\textwidth]{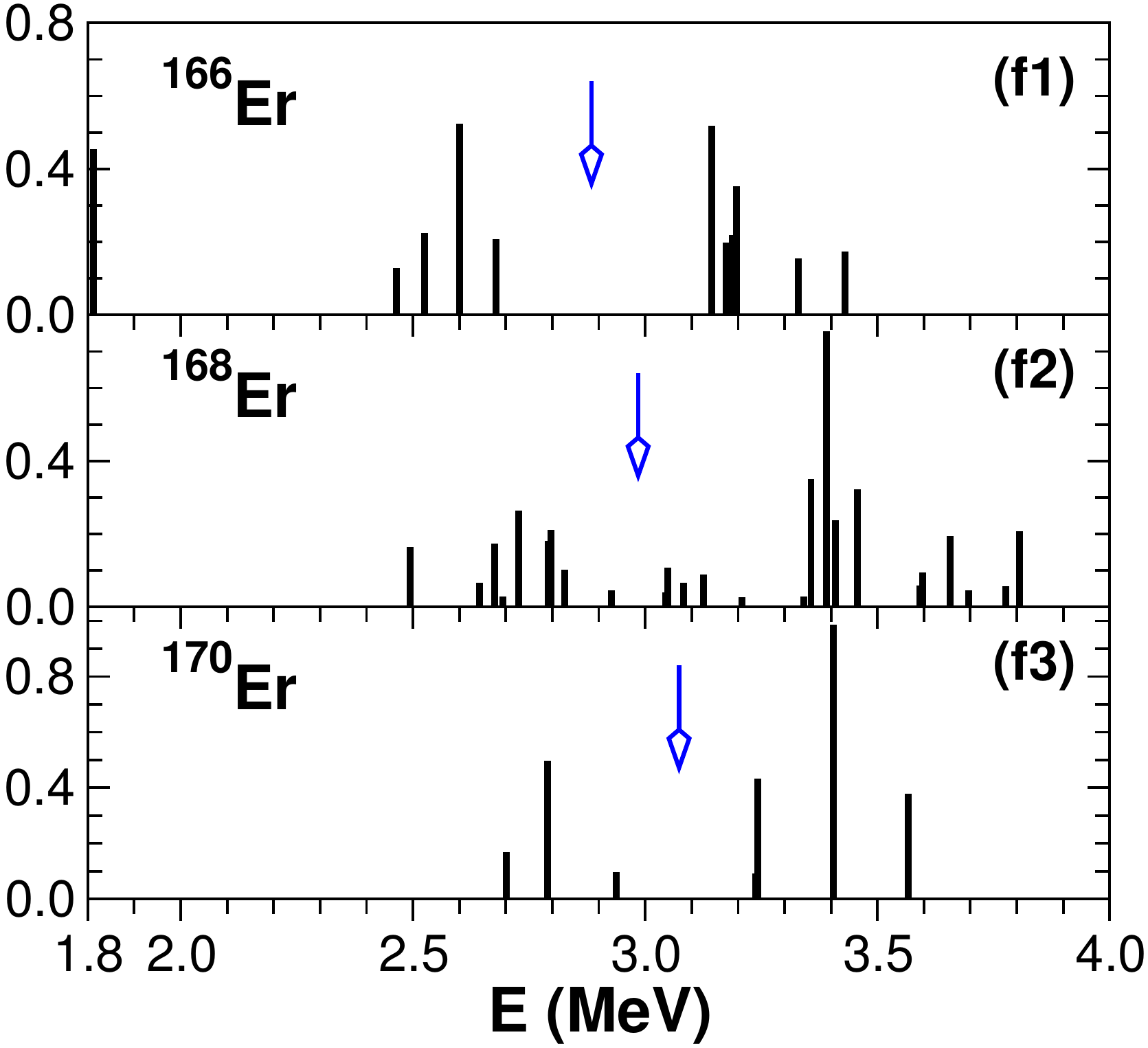}\hspace*{1mm}\includegraphics[width=0.305\textwidth]{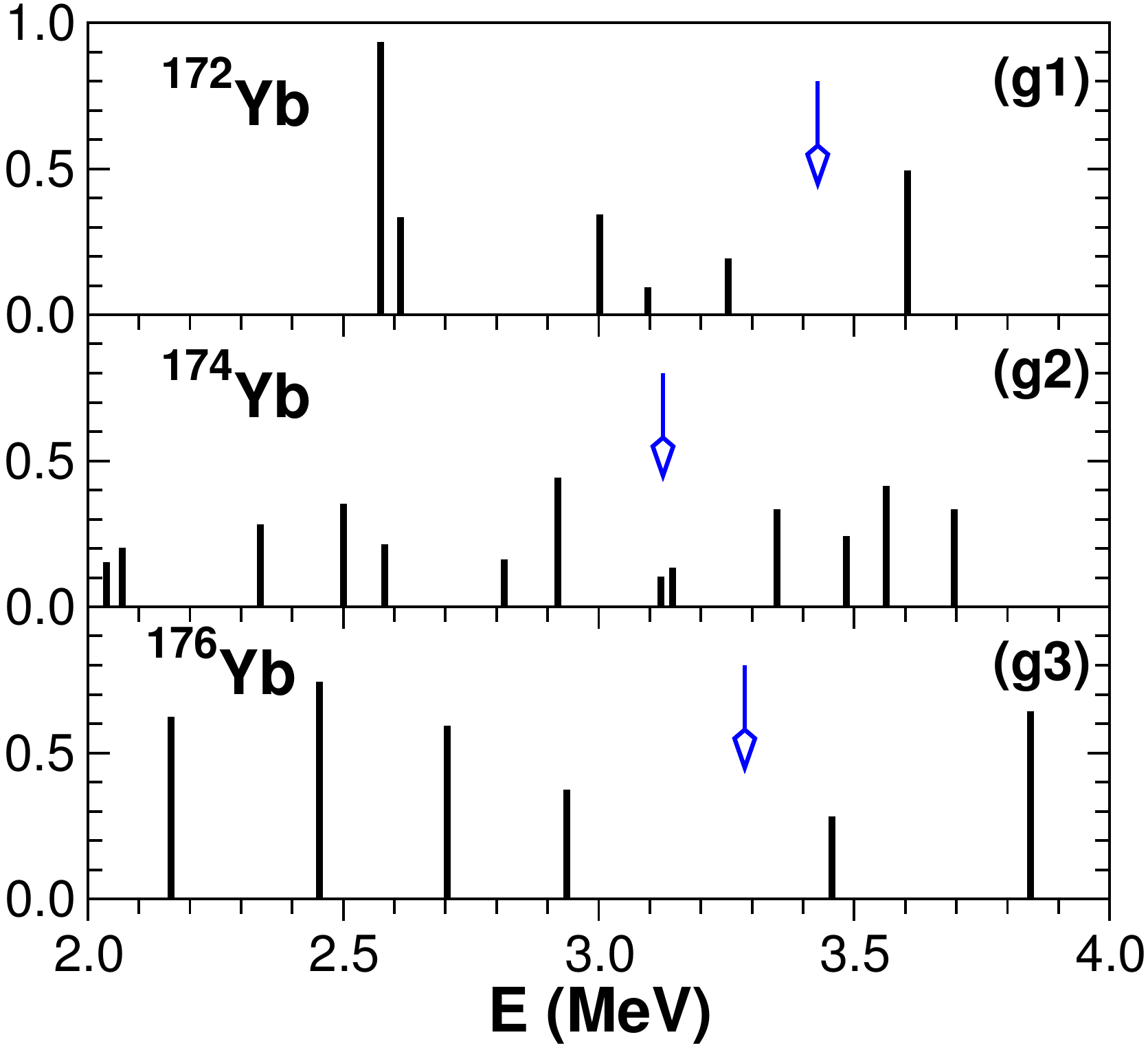}\vspace*{1mm}
\includegraphics[width=0.333\textwidth]{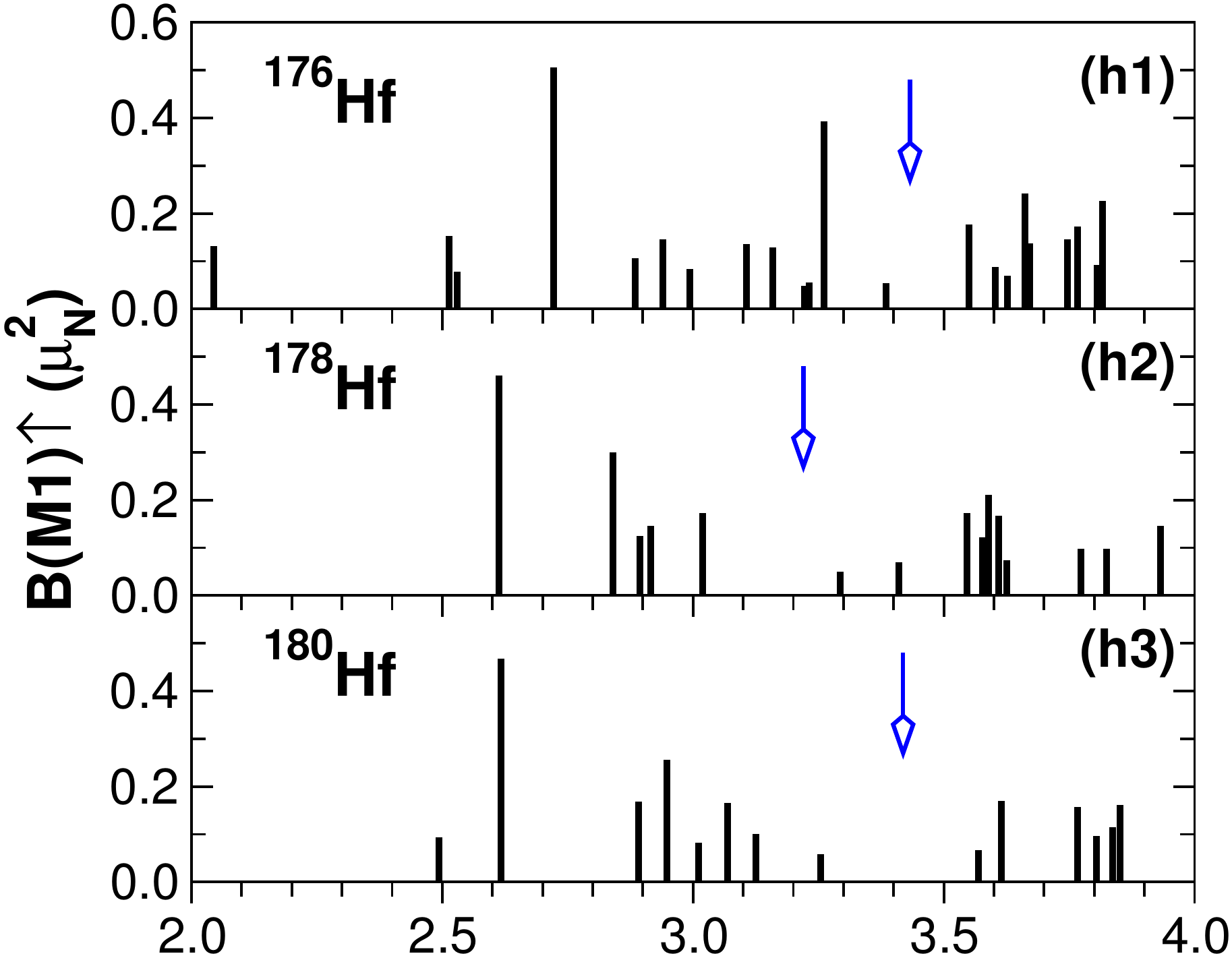}\hspace*{1mm}\includegraphics[width=0.305\textwidth]{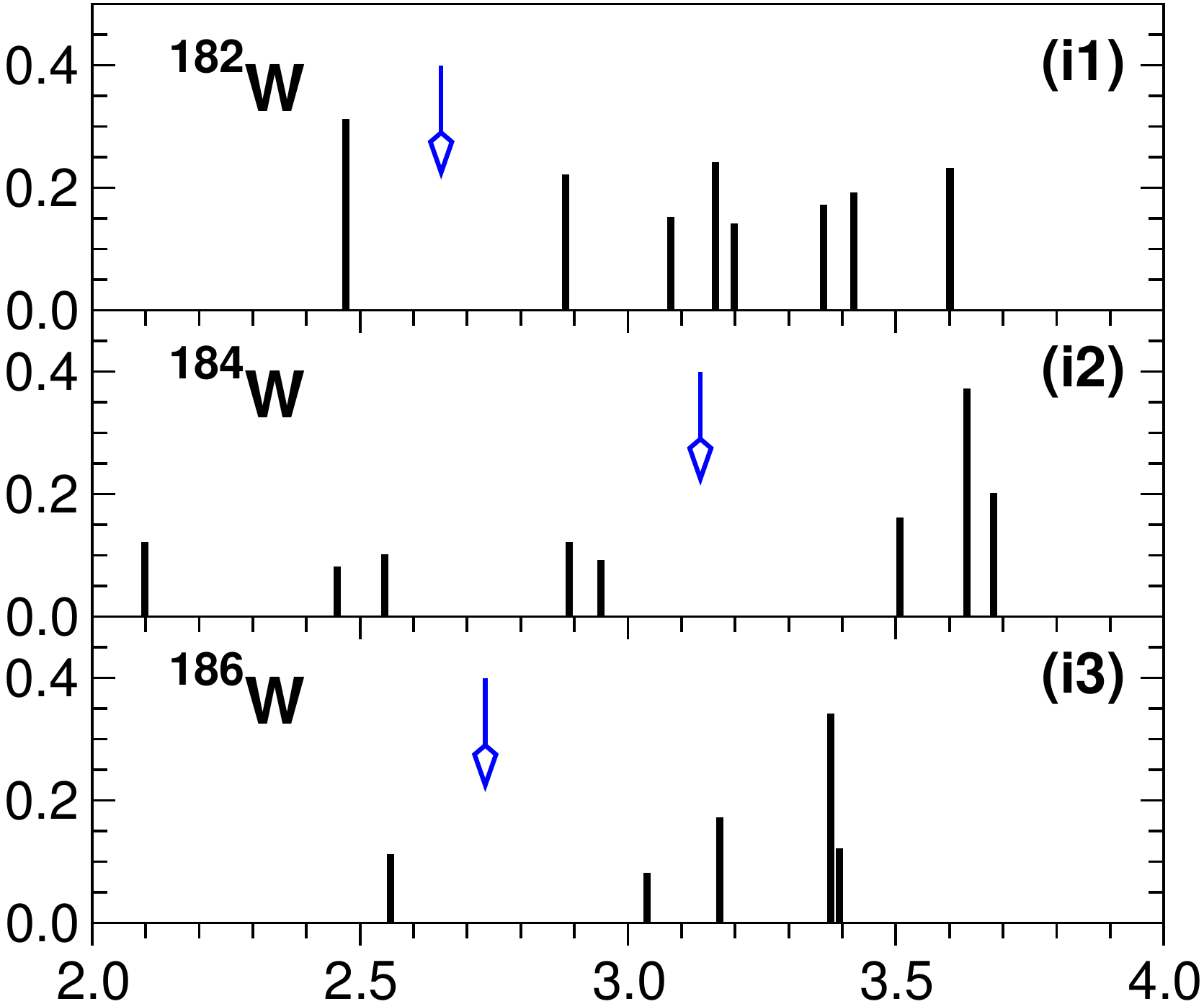}
\includegraphics[width=0.333\textwidth]{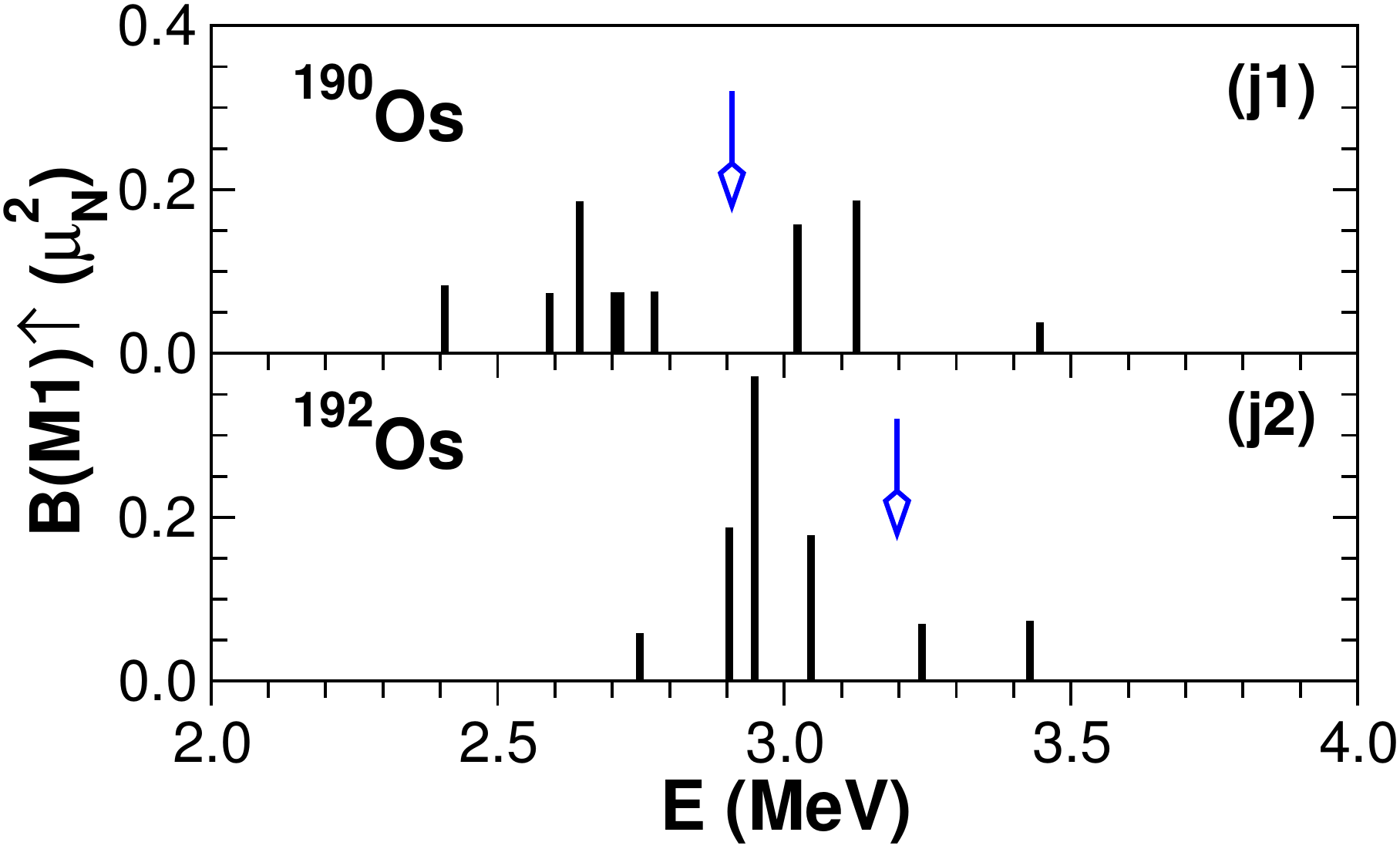}\hspace*{1mm}\includegraphics[width=0.305\textwidth]{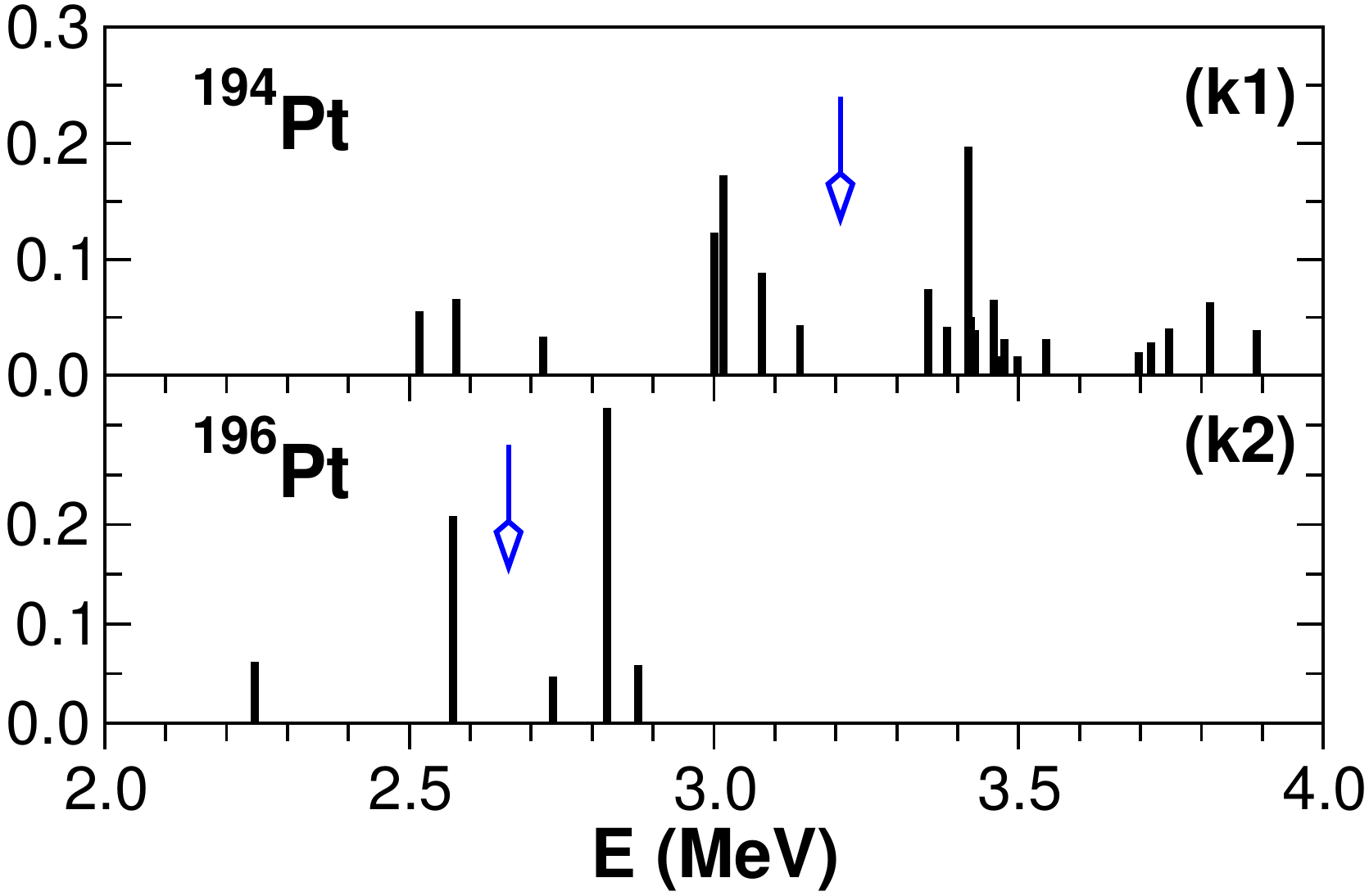}
\caption{(Color online) The experimentally observed spectra of $1^+$ excitations: $^{134}$Ba -- \cite{Ba},
$^{144-150}$Nd -- \cite{Nd144,Margr}, $^{148-154}$Sm -- \cite{Sm}, $^{154-160}$Gd -- \cite{Gd,Gd160}, 
$^{160-164}$Dy -- \cite{Wessel,Dy}, $^{166-170}$Er -- \cite{Er}, $^{172-176}$Yb -- \cite{Yb}, 
$^{176-180}$Hf -- \cite{Hf176,Hf}, $^{182-186}$W  -- \cite{W},  $^{190-192}$Os -- \cite{Os},
$^{194-196}$Pt -- \cite{Pt194,Pt196}. 
The point of the division of the spectrum in two parts is shown by a blue arrow.
}
\label{fig4}
\end{figure*}

\begin{table*}[h!]\vspace*{-5mm}
\caption{\label{tab3} Energy centroids $E$ and summed
transition probabilities $B(M1)$ of the spin and orbital scissors; spectra are taken 
from papers shown in the last column. The information for many 
nuclei contains second line, which is obtained by taking into account $1^{+}$ 
states together with $1^{\pi}$ states whith $\pi$ unknown. }
\begin{ruledtabular}
\begin{tabular}{lcccccccccclcc}
 Nuclei & 
\multicolumn{6}{c}{ $E$ (MeV)}    &
\multicolumn{6}{c}{ $B(M1)\ (\mu_N^2)$}   & Ref.      \\
\cline{2-7} \cline{8-13}
 &\multicolumn{2}{c}{spin scissors} &\multicolumn{2}{c}{orbital scissors}  &\multicolumn{2}{c}{centroid} & 
  \multicolumn{2}{c}{spin scissors} &\multicolumn{2}{c}{orbital scissors}  &\multicolumn{2}{c}{ $\sum$ } &  \\
\cline{2-3} \cline{4-5} \cline{6-7} \cline{8-9} \cline{10-11} \cline{12-13}  
 & Exp. & WFM & Exp. & WFM & Exp. & WFM & 
   Exp. & WFM & Exp. & WFM &\multicolumn{1}{c}{Exp.} & WFM & \\
\hline   
 $^{134}$Ba  & 2.88 & 3.02 & 3.35 & 3.43 & 2.99 & 3.04 & 0.43(06) & 0.65 & 0.13(03) & 0.03 & 0.56(09)   & 0.68 & \cite{Ba}    \\
             & 2.85 &      & 3.65 &      & 3.33 &      & 0.51(08) &      & 0.76(16) &      & 1.26(24)   &      &     \\ 
 $^{144}$Nd  & 2.57 & 2.97 & 3.57 & 3.32 & 3.07 & 3.21 & 0.39(03) & 0.06 & 0.39(02) & 0.14 & 0.78(05)   & 0.20 & \cite{Nd144} \\
             & 2.63 &      & 3.64 &      & 3.22 &      & 0.51(05) &      & 0.70(06) &      & 1.21(11)   &      &     \\                         
 $^{146}$Nd  & 2.36 & 3.00 & 3.37 & 3.38 & 2.90 & 3.20 & 0.34(04) & 0.27 & 0.39(06) & 0.30 & 0.73(10)   & 0.57 & \cite{Margr} \\
             & 2.38 &      & 3.60 &      & 3.28 &      & 0.38(05) &      & 1.08(18) &      & 1.45(23)   &      &     \\                     
 $^{148}$Nd  & 3.22 & 3.09 & 3.83 & 3.54 & 3.40 & 3.22 & 0.80(19) & 0.91 & 0.32(07) & 0.37 & 1.12(26)   & 1.28 & \cite{Margr} \\ 
 $^{150}$Nd  & 3.02 & 2.90 & 3.72 & 3.64 & 3.12 & 3.13 & 1.56(21) & 1.27 & 0.27(05) & 0.56 & 1.83(26)   & 1.83 & \cite{Margr} \\ 
 $^{148}$Sm  &  --  & 2.98 &  --  & 3.35 & 3.07 & 3.17 &  --      & 0.22 &  --      & 0.24 & 0.51(12)   & 0.46 & \cite{Sm}    \\
 $^{150}$Sm  & 3.07 & 3.06 & 3.73 & 3.49 & 3.18 & 3.17 & 0.81(12) & 0.84 & 0.16(05) & 0.27 & 0.97(17)   & 1.12 & \cite{Sm}    \\             
 $^{152}$Sm  &  --  & 2.71 &  --  & 3.53 & 2.97 & 2.99 &  --      & 1.65 &  --      & 0.85 & 2.41(33)   & 2.50 & \cite{Sm}    \\
 $^{154}$Sm  & 2.98 & 2.79 & 3.62 & 3.64 & 3.14 & 3.10 & 2.08(30) & 2.12 & 0.68(20) & 1.22 & 2.76(50)   & 3.34 & \cite{Sm}    \\     
 $^{154}$Gd  & 2.91 & 2.75 & 3.10 & 3.58 & 3.00 & 3.04 & 1.40(25) & 1.95 & 1.20(25) & 1.05 & 2.60(50)   & 3.00 &  \cite{Gd}    \\               
 $^{156}$Gd  & 2.28 & 2.79 & 3.06 & 3.63 & 2.94 & 3.09 & 0.49(12) & 2.19 & 2.73(56) & 1.24 & 3.22(68)   & 3.44 &  \cite{Gd}    \\                 
 $^{158}$Gd  & 2.64 & 2.78 & 3.20 & 3.62 & 3.04 & 3.09 & 1.13(23) & 2.25 & 2.86(42) & 1.27 & 3.99(65)   & 3.52 &  \cite{Gd}    \\                   
 $^{160}$Gd  & 2.65 & 2.82 & 3.33 & 3.67 & 3.10 & 3.14 & 1.53(14) & 2.53 & 2.88(40) & 1.49 & 4.41(54)   & 4.02 &  \cite{Gd160} \\
 $^{160}$Dy  & 2.84 & 2.78 & 3.06 & 3.61 & 2.87 & 3.08 & 2.12(25) & 2.30 & 0.30(05) & 1.30 & 2.42(30)   & 3.60 & \cite{Wessel} \\
 $^{162}$Dy  & 2.44 & 2.78 & 2.96 & 3.60 & 2.84 & 3.07 & 0.71(05) & 2.37 & 2.59(19) & 1.32 & 3.30(24)   & 3.69 & \cite{Dy}     \\   
 $^{164}$Dy  & --   & 2.77 & --   & 3.60 & 3.17 & 3.07 & --       & 2.44 & --       & 1.36 & 3.85(31)   & 3.80 & \cite{Dy}     \\   
 $^{166}$Er  & 2.36 & 2.77 & 3.21 & 3.59 & 2.79 & 3.06 & 1.52(34) & 2.48 & 1.60(24) & 1.38 & 3.12(58)   & 3.86 & \cite{Er}     \\
             & 2.37 &      & 3.25 &      & 2.85 &      & 1.56(35) &      & 1.86(46) &      & 3.42(81)   &      &     \\
 $^{168}$Er  & 2.72 & 2.77 & 3.43 & 3.58 & 3.21 & 3.06 & 1.21(14) & 2.54 & 2.63(35) & 1.41 & 3.85(50)   & 3.95 & \cite{Er}     \\
             & 2.69 &      & 3.46 &      & 3.22 &      & 1.35(17) &      & 3.04(44) &      & 4.38(61)   &      &     \\             
 $^{170}$Er  & 2.79 & 2.76 & 3.39 & 3.57 & 3.22 & 3.05 & 0.75(11) & 2.60 & 1.88(28) & 1.43 & 2.63(39)   & 4.03 & \cite{Er}     \\
             & 2.77 &      & 3.42 &      & 3.21 &      & 1.07(19) &      & 2.24(41) &      & 3.30(59)   &      &     \\         
 $^{172}$Yb  & 2.75 & 2.72 & 3.60 & 3.51 & 2.93 & 2.99 & 1.88(37) & 2.46 & 0.49(12) & 1.26 & 2.37(49)   & 3.72 & \cite{Yb}     \\
             & 2.77 &      & 3.73 &      & 3.07 &      & 1.99(41) &      & 0.94(26) &      & 2.93(67)   &      &     \\
 $^{174}$Yb  & 2.56 & 2.72 & 3.49 & 3.50 & 2.96 & 2.98 & 1.89(70) & 2.51 & 1.44(51) & 1.29 & 3.33(1.21) & 3.80 & \cite{Yb}     \\
             & 2.35 &      & 3.29 &      & 2.97 &      & 1.19(52) &      & 2.28(76) &      & 3.47(1.28) &      &     \\
 $^{176}$Yb  & 2.52 & 2.68 & 3.73 & 3.44 & 2.86 & 2.92 & 2.32(60) & 2.34 & 0.92(45) & 1.12 & 3.24(1.05) & 3.46 & \cite{Yb}     \\     
             & 2.52 &      & 3.65 &      & 2.98 &      & 2.32(60) &      & 1.60(70) &      & 3.92(1.30) &      &     \\     
 $^{176}$Hf  & 2.89 & 3.06 & 3.70 & 3.73 & 3.22 & 3.27 & 1.99(15) & 2.15 & 1.33(13) & 1.00 & 3.32(28)   & 3.15 & \cite{Hf176}  \\
             & 2.91 &      & 3.69 &      & 3.25 &      & 2.47(22) &      & 1.93(23) &      & 4.40(45)   &      &     \\
 $^{178}$Hf  & 2.79 & 3.05 & 3.64 & 3.72 & 3.21 & 3.27 & 1.19(11) & 2.18 & 1.19(22) & 1.02 & 2.38(33)   & 3.20 & \cite{Hf}     \\
             & 2.73 &      & 3.67 &      & 3.15 &      & 1.19(11) &      & 1.38(26) &      & 2.57(37)   &      &     \\
 $^{180}$Hf  & 2.84 & 3.02 & 3.75 & 3.67 & 3.16 & 3.22 & 1.38(17) & 1.97 & 0.75(13) & 0.86 & 2.13(30)   & 2.84 & \cite{Hf}     \\
             & 2.84 &      & 3.80 &      & 3.30 &      & 1.57(22) &      & 1.43(27) &      & 3.00(49)   &      &     \\     
 $^{182}$W   & 2.47 & 2.96 & 3.25 & 3.57 & 3.10 & 3.13 & 0.31(05) & 1.48 & 1.34(23) & 0.55 & 1.65(28)   & 2.03 & \cite{W}    \\
 $^{184}$W   & 2.58 & 2.94 & 3.62 & 3.52 & 3.19 & 3.08 & 0.51(10) & 1.27 & 0.73(27) & 0.41 & 1.24(37)   & 1.68 & \cite{W}    \\
 $^{186}$W   & 2.56 & 2.91 & 3.29 & 3.48 & 3.19 & 3.03 & 0.11(01) & 1.06 & 0.71(20) & 0.29 & 0.82(21)   & 1.36 & \cite{W}    \\
             & 2.56 &      & 3.36 &      & 3.29 &      & 0.11(01) &      & 1.18(64) &      & 1.29(65)   &      &     \\         
%
 $^{190}$Os  & 2.64 & 2.87 & 3.11 & 3.25 & 2.83 & 2.98 & 0.56(08) & 1.01 & 0.38(04) & 0.38 & 0.94(12)   & 1.39 & \cite{Os}    \\
             & 2.14 &      & 3.37 &      & 2.72 &      & 1.26(19) &      & 1.12(16) &      & 2.38(36)   &      &     \\
 $^{192}$Os  & 2.95 & 2.85 & 3.34 & 3.21 & 3.00 & 2.94 & 0.79(04) & 0.74 & 0.14(02) & 0.25 & 0.93(06)   & 0.99 & \cite{Os}    \\     
             & 2.89 &      & 3.40 &      & 3.07 &      & 1.16(08) &      & 0.63(06) &      & 1.78(14)   &      &     \\ 
 $^{194}$Pt  & 2.92 & 2.82 & 3.52 & 3.16 & 3.25 & 2.97 & 0.57(09) & 0.66 & 0.74(14) & 0.52 & 1.31(23)   & 1.17 & \cite{Pt194} \\
             & 2.92 &      & 3.51 &      & 3.29 &      & 0.57(09) &      & 0.93(17) &      & 1.50(25)   &      &     \\ 
 $^{196}$Pt  & 2.50 & 2.81 & 2.82 & 3.12 & 2.70 & 2.95 & 0.27(05) & 0.46 & 0.42(07) & 0.40 & 0.69(13)   & 0.86 & \cite{Pt196} \\   
             & 2.50 &      & 2.93 &      & 2.79 &      & 0.27(05) &      & 0.55(15) &      & 0.82(20)   &      &     \\ 
\end{tabular}                                                                                                         
\end{ruledtabular}
\end{table*} 

\begin{figure}[h!]
\includegraphics[width=\columnwidth]{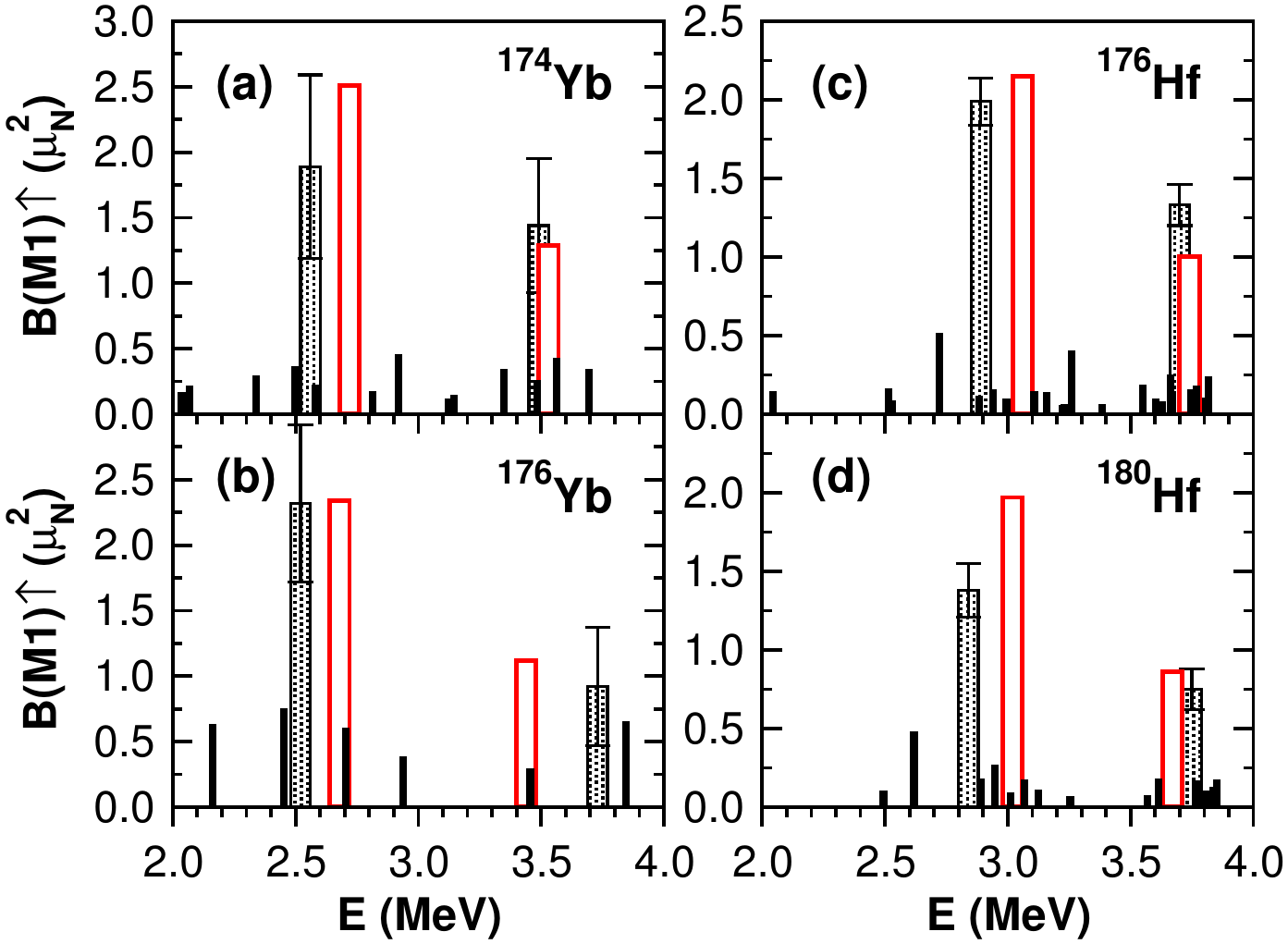}
\caption{(Color online) 
The centroids of experimentally observed spectra of $1^+$ excitations in 
$^{174,176}$Yb (a, b) and $^{176,180}$Hf (c, d) (black rectangles with error bars) are 
compared with the results of calculations (red rectangles) 
of the spin and orbital scissors modes. Notice that the strength of the lower peak (spin-scissors) is always stronger than the one of the upper peak 
(orbital-scissors).}
\label{WFM_1}
\end{figure}

\section{Experimental situation in rare earths}\label{s3}

We here will perform a 
systematic analysis of experimental data for rare earth nuclei, where
the majority of nuclear scissors are found. 
We have studied practically all papers containing experimental data for low 
lying $M1$ excitations in rare earth nuclei. For the sake of convenience we have 
collected in Fig.~\ref{fig4} 
all experimentally known spectra of low lying $1^+$ excitations \cite{Wessel,Ba,Nd144,Margr,Sm,Gd,Gd160,Dy,Er,Yb,Hf176,Hf,W,Os,Pt194,Pt196}. It should be 
emphasized, that only the levels with positive parities are displayed here. As a
matter of fact, there exist many $1^{\pi}$ excitations in the energy interval
1~MeV$ - 4$~MeV, whose parities $\pi$ are not known up to now. 

\begin{figure}[b!]
\includegraphics[width=\columnwidth]{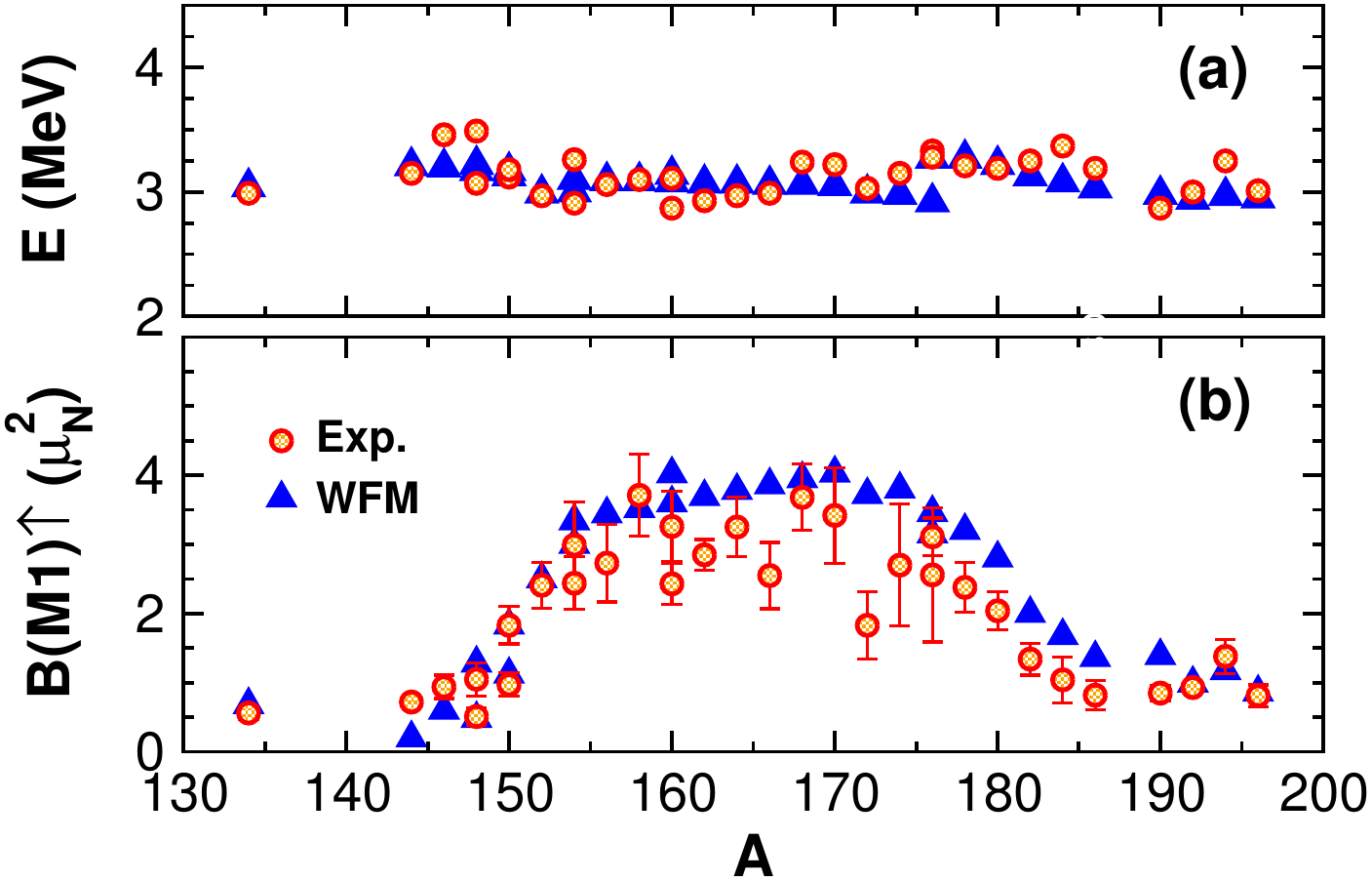}
\caption{(Color online) 
Calculated (WFM) mean excitation energies (a) and  summed $M1$ strengths (b) of 
the scissors mode are compared with experimental data (Exp.) from~\cite{Rich}.}\label{fig8}
\end{figure}

One glance on Fig.~\ref{fig4} is
enough to understand that the situation with spectra in rare earth nuclei iscomplicated.
Nevertheless, we will try to proceed in the same way as with the Actinides: we fold the spectra with a Lorentzian whose width is increased 
until only two humps survive. 
The minimum between the two humps is indicated with an arrow in Fig.~\ref{fig4}.
The states in the low lying group are then identified as belonging to the spin scissors motion and the ones of high lying group as orbital scissors states.
The values of $B(M1)$ and energy centroids, corresponding to the described separation, 
 are compared with theoretical results in the Table~\ref{tab3}.
 Studying attentively the
Table~\ref{tab3} one can 
find satisfactory agreement between theoretical and experimental results for 
  the orbital scissors in 13 nuclei:
$^{146,148}$Nd, $^{154}$Gd, $^{166,170}$Er,
$^{174,176}$Yb, $^{176,178,180}$Hf, $^{190}$Os and $^{194,196}$Pt. 
  The same degree of agreement can be found for the spin scissors in 15 nuclei:
$^{134}$Ba, $^{146,148,150}$Nd, $^{150,154}$Sm, $^{154}$Gd, $^{160}$Dy, $^{172,174,176}$Yb, $^{176,180}$Hf, $^{192}$Os and $^{194}$Pt. 
Therefore the satisfactory agreement between theoretical and experimental 
results for both, the orbital and spin scissors, is observed in 8 nuclei: 
$^{146,148}$Nd, $^{154}$Gd, $^{174,176}$Yb, $^{176,180}$Hf and $^{194}$Pt.
Again, as in Fig.~\ref{WFM_00} for the Actinides, in Fig.~\ref{WFM_1} we compare for four rare earth nuclei the experimental centroids with our results. 
The agreement is nearly perfect. In the other four nuclei the agreement is less good but still acceptable. 
The harvest seems to be rather meager. However, we have to remember that spin and orbital scissors are surely mostly not quite separated. 
It is a lucky accident when they are clearly separated like in the Th and (eventually) in Pa isotopes and here as in Fig.~\ref{WFM_1} .


Therefore, an additional handful of nuclei in the rare earth region is a very welcome support of our theoretical 
analysis of the existence of two separate sissors modes: spin and orbital. This the more so as we get at least 
semi-quantitative agreement between experiment and theory, in the sense that for all those selected nuclei the
transition probability for spin scissors is stronger than for orbital scissors. 
This finding gives support to our theoretical analysis.
It is
worth noting at the end of this section, that sums of experimental $B(M1)$ values and 
the respective energy centroids agree (with rare exceptions) very well with 
theoretical predictions (see Fig.~\ref{fig8}).

\section{Conclusions and Outlook}\label{s7}

The aim of this paper was to find experimental indications about the existence of
the "spin" scissors mode predicted in our previous publications 
\cite{BaMo,BaMoPRC,BaMoPRC2}. To this end we have performed a detailed analysis of 
experimental data on $1^+$ excitations for actinides and for Rare Earth nuclei of the $N=82-126$ major shell. 

First of all let us again comment on the very clear experimental situation in the actinides 
for heated nuclei of Th, Pa, U isotopes. In all cases a clear two humps structure has been revealed,  
see Figs.~\ref{figSiem} and~\ref{U}. Unfortunately with our WFM technique, we are so far 
not prepared to investigate nuclei at finite temperature. 
So we tried in this work to investigate scissors modes on top of the ground states. 
We tried to be as exhaustive in the presentation of published data in rare earth nuclei and the 
actinides as possible. For the actinides, there exists a very clear cut example given by $^{232}$Th, 
see Figs.~\ref{figAdekola} and~\ref{fig5}, and our results are in good agreement with the experimental 
values for position and $B(M1)$ values. 
We stress again that this concerns also the feature that the $B(M1)$'s are stronger for the hypothetical 
spin scissors than for the orbital one. For the Uranium isotopes the separation into two peaks of the 
$M1$ excitations is not so clear. So we tried to find signatures of splitting also in the rare earth nuclei. 
Despite of the fact that one can imagine that spin-scissors  and orbital ones are not well 
separated in nuclei, lighter than the actinides, we nevertheless found a handful of  
examples where our method of  separation into a high lying and low lying group works and yields satisfying agreement with experiment. 
Here we again found that 
$B(M1)$'s are stronger for low lying than for high lying part of levels. So, there are about half a dozen 
examples which support our theoretical findings that there exist two groups of scissors modes, 
the spin scissors and orbital scissors modes.
Actually we divided the experimental spectra of almost all rare earth nuclei shown in Fig.~\ref{fig4} 
into low lying and higher lying parts indicated by the blue arrows. Without giving quantitative agreement with experiment besides for those 8 nuclei mentioned 
in the main text, we at least found most of the 
time  that the low lying parts have stronger $B(M1)$'s than high lying groups again in qualitative agreement 
with our theoretical investigation. In conclusion, there seems to exist as well some support of the rare earth nuclei for the existence of the spin scissors mode. 
Let us mention again that we obtained the two hump structures in folding the spectra with a Lorentzian (we also could have taken as well a Gaussian). 
This method of separating the often quite complex splitting patterns of the $M1$ strength in just two humps is an assumption that derives from the present 
theoretical description where besides the orbital scissors part, a spin-scissors part appears.
A special mention may need the clear situation in $^{232}$Th because there  exists a QRPA 
calculation~\cite{Kuliev}. This QRPA calculation also reveals a two hump structure in close agreement 
with experiment and with our findings. 
We may finish off with the remark that it does not seem evident to disentangle a QRPA spectrum into spin and orbital scissors states. 
For this it will be necessary to find the excitation operators which specifically excite the spin and orbital scissors individually. 
For the moment this task can only be achieved with our WFM method
because the amplitudes have a clear connection to the physics of the obtained states. 
It will be a future project to make detailed comparison between the QRPA and our calculations. 
A further aim is to introduce finite temperature into the WFM formalism what eventually would allow to explain the strongly enhanced $B(M1)$ transitions found 
experimentally in the Actinides.

\begin{acknowledgments}
We wish to thank A. Richter for valuable remarks, M. Guttormsen and J. N. Wilson for fruitful discussions, A. A. Kuliev for sharing the results of his calculations 
and P. von Neuman-Cosel who attracted our attention to the paper of
\mbox{A. S. Adekola {\it et al} \cite{Adekola}}.
The work was supported by the \mbox{IN2P3/CNRS-JINR 03-57} Collaboration agreement. 
\end{acknowledgments}

\appendix

\section{WFM method and details of calculations}
\label{AppA}

The basis of our method is the  Time-Dependent Hartree--Fock--Bogoliubov (TDHFB)
equation in matrix formulation~\cite{Ring}:
\begin{equation}
i\hbar\dot\R=[\H,\R]
\label{tHFB}
\end{equation}
with
\begin{equation}
\R={\hat\rho\qquad-\hat\kappa\choose-\hat\kappa^{\dagger}\;\;1-\hat\rho^*},
\quad\H={\hat
h\quad\;\;\hat\Delta\choose\hat\Delta^{\dagger}\quad-\hat h^*}
\end{equation}
The normal density matrix $\hat \rho$ and Hamiltonian $\hat h$ are
hermitian whereas the abnormal density $\hat\kappa$ and the pairing
gap $\hat\Delta$ are skew symmetric: $\hat\kappa^{\dagger}=-\hat\kappa^*$, 
\mbox{$\hat\Delta^{\dagger}=-\hat\Delta^*$}.

The detailed form of the TDHFB equations is
\begin{eqnarray}
&& i\hbar\dot{\hat\rho} =\hat h\hat\rho -\hat\rho\hat h
-\hat\Delta \hat\kappa ^{\dagger}+\hat\kappa \hat\Delta^\dagger,
\nonumber\\
&&-i\hbar\dot{\hat\rho}^*=\hat h^*\hat\rho ^*-\hat\rho ^*\hat h^*
-\hat\Delta^\dagger\hat\kappa +\hat\kappa^\dagger\hat\Delta ,
\nonumber\\
&&-i\hbar\dot{\hat\kappa} =-\hat h\hat\kappa -\hat\kappa \hat h^*+\hat\Delta
-\hat\Delta \hat\rho ^*-\hat\rho \hat\Delta ,
\nonumber\\
&&-i\hbar\dot{\hat\kappa}^\dagger=\hat h^*\hat\kappa^\dagger
+\hat\kappa^\dagger\hat h-\hat\Delta^\dagger
+\hat\Delta^\dagger\hat\rho +\hat\rho^*\hat\Delta^\dagger.\qquad
\label{HFB}
\end{eqnarray}
 We do not specify the isospin indices in order to make
formulae more transparent. Let us consider matrix form of (\ref{HFB}) in coordinate 
space keeping spin indices $s, s'$ with
compact notation \mbox{$X_{rr'}^{ss'}\equiv\langle \br,s|\hat X|\br',s'\rangle $}. Then
the set of TDHFB equations with specified spin indices reads~\cite{BaMoPRC2}:
\begin{widetext}
\begin{eqnarray}
&&i\hbar\dot{\rho}_{rr''}^{\uparrow\uparrow} =
\int\!d^3r'(
 h_{rr'}^{\uparrow\uparrow}\rho_{r'r''}^{\uparrow\uparrow} 
-\rho_{rr'}^{\uparrow\uparrow} h_{r'r''}^{\uparrow\uparrow}
+\hat h_{rr'}^{\uparrow\downarrow}\rho_{r'r''}^{\downarrow\uparrow} 
-\rho_{rr'}^{\uparrow\downarrow} h_{r'r''}^{\downarrow\uparrow}
-\Delta_{rr'}^{\uparrow\downarrow}{\kappa^{\dagger}}_{r'r''}^{\downarrow\uparrow}
+\kappa_{rr'}^{\uparrow\downarrow}{\Delta^{\dagger}}_{r'r''}^{\downarrow\uparrow}),
\nonumber\\
&&i\hbar\dot{\rho}_{rr''}^{\uparrow\downarrow} =
\int\!d^3r'(
 h_{rr'}^{\uparrow\uparrow}\rho_{r'r''}^{\uparrow\downarrow} 
-\rho_{rr'}^{\uparrow\uparrow} h_{r'r''}^{\uparrow\downarrow}
+\hat h_{rr'}^{\uparrow\downarrow}\rho_{r'r''}^{\downarrow\downarrow} 
-\rho_{rr'}^{\uparrow\downarrow} h_{r'r''}^{\downarrow\downarrow}),
\nonumber\\
&&i\hbar\dot{\rho}_{rr''}^{\downarrow\uparrow} =
\int\!d^3r'(
 h_{rr'}^{\downarrow\uparrow}\rho_{r'r''}^{\uparrow\uparrow} 
-\rho_{rr'}^{\downarrow\uparrow} h_{r'r''}^{\uparrow\uparrow}
+\hat h_{rr'}^{\downarrow\downarrow}\rho_{r'r''}^{\downarrow\uparrow} 
-\rho_{rr'}^{\downarrow\downarrow} h_{r'r''}^{\downarrow\uparrow}),
\nonumber\\
&&i\hbar\dot{\rho}_{rr''}^{\downarrow\downarrow} =
\int\!d^3r'(
 h_{rr'}^{\downarrow\uparrow}\rho_{r'r''}^{\uparrow\downarrow} 
-\rho_{rr'}^{\downarrow\uparrow} h_{r'r''}^{\uparrow\downarrow}
+\hat h_{rr'}^{\downarrow\downarrow}\rho_{r'r''}^{\downarrow\downarrow} 
-\rho_{rr'}^{\downarrow\downarrow} h_{r'r''}^{\downarrow\downarrow}
-\Delta_{rr'}^{\downarrow\uparrow}{\kappa^{\dagger}}_{r'r''}^{\uparrow\downarrow}
+\kappa_{rr'}^{\downarrow\uparrow}{\Delta^{\dagger}}_{r'r''}^{\uparrow\downarrow}),
\nonumber\\
&&i\hbar\dot{\kappa}_{rr''}^{\uparrow\downarrow} = -\hat\Delta_{rr''}^{\uparrow\downarrow}
+\int\!d^3r'\left(
 h_{rr'}^{\uparrow\uparrow}\kappa_{r'r''}^{\uparrow\downarrow} 
+\kappa_{rr'}^{\uparrow\downarrow} {h^*}_{r'r''}^{\downarrow\downarrow}
+\Delta_{rr'}^{\uparrow\downarrow}{\rho^*}_{r'r''}^{\downarrow\downarrow} 
+\rho_{rr'}^{\uparrow\uparrow}\Delta_{r'r''}^{\uparrow\downarrow}
\right),
\nonumber\\
&&i\hbar\dot{\kappa}_{rr''}^{\downarrow\uparrow} = -\hat\Delta_{rr''}^{\downarrow\uparrow}
+\int\!d^3r'\left(
 h_{rr'}^{\downarrow\downarrow}\kappa_{r'r''}^{\downarrow\uparrow} 
+\kappa_{rr'}^{\downarrow\uparrow} {h^*}_{r'r''}^{\uparrow\uparrow}
+\Delta_{rr'}^{\downarrow\uparrow}{\rho^*}_{r'r''}^{\uparrow\uparrow} 
+\rho_{rr'}^{\downarrow\downarrow}\Delta_{r'r''}^{\downarrow\uparrow}
\right).
\label{HFsp}
\end{eqnarray}
This set of equations must be complemented by the complex conjugated equations.

We work with the Wigner transform \cite{Ring} of 
equations (\ref{HFsp}). The relevant mathematical details can be found in
\cite{BaMoPRC2,Malov}. 
We do not write out the coordinate 
dependence~$(\br,\bp)$ of all functions in order to make the formulae 
more transparent. We have
\begin{eqnarray}
      i\hbar\dot f^{\uparrow\uparrow} &=&i\hbar\{h^{\uparrow\uparrow},f^{\uparrow\uparrow}\}
+h^{\uparrow\downarrow}f^{\downarrow\uparrow}-f^{\uparrow\downarrow}h^{\downarrow\uparrow}
+\frac{i\hbar}{2}\{h^{\uparrow\downarrow},f^{\downarrow\uparrow}\}
-\frac{i\hbar}{2}\{f^{\uparrow\downarrow},h^{\downarrow\uparrow}\}
\nonumber\\
&-&\frac{\hbar^2}{8}\{\!\{h^{\uparrow\downarrow},f^{\downarrow\uparrow}\}\!\}
+\frac{\hbar^2}{8}\{\!\{f^{\uparrow\downarrow},h^{\downarrow\uparrow}\}\!\} 
+ \kappa\Delta^* - \Delta\kappa^* 
\nonumber
\\
&+&\frac{i\hbar}{2}\{\kappa,\Delta^*\}-\frac{i\hbar}{2}\{\Delta,\kappa^*\}
- \frac{\hbar^2}{8}\{\!\{\kappa,\Delta^*\}\!\} + \frac{\hbar^2}{8}\{\!\{\Delta,\kappa^*\}\!\}
+...,
\nonumber\\
      i\hbar\dot f^{\downarrow\downarrow} &=&i\hbar\{h^{\downarrow\downarrow},f^{\downarrow\downarrow}\}
+h^{\downarrow\uparrow}f^{\uparrow\downarrow}-f^{\downarrow\uparrow}h^{\uparrow\downarrow}
+\frac{i\hbar}{2}\{h^{\downarrow\uparrow},f^{\uparrow\downarrow}\}
-\frac{i\hbar}{2}\{f^{\downarrow\uparrow},h^{\uparrow\downarrow}\}
\nonumber\\
&-&\frac{\hbar^2}{8}\{\!\{h^{\downarrow\uparrow},f^{\uparrow\downarrow}\}\!\}
+\frac{\hbar^2}{8}\{\!\{f^{\downarrow\uparrow},h^{\uparrow\downarrow}\}\!\} 
+ \bar\Delta^* \bar\kappa - \bar\kappa^* \bar\Delta
\nonumber\\
&+&\frac{i\hbar}{2}\{\bar\Delta^*,\bar\kappa\}-\frac{i\hbar}{2}\{\bar\kappa^*,\bar\Delta\}
- \frac{\hbar^2}{8}\{\!\{\bar\Delta^*,\bar\kappa\}\!\} + \frac{\hbar^2}{8}\{\!\{\bar\kappa^*,\bar\Delta\}\!\}
+...,
\nonumber\\
      i\hbar\dot f^{\uparrow\downarrow} &=&
f^{\uparrow\downarrow}(h^{\uparrow\uparrow}-h^{\downarrow\downarrow})
+\frac{i\hbar}{2}\{(h^{\uparrow\uparrow}+h^{\downarrow\downarrow}),f^{\uparrow\downarrow}\}
-\frac{\hbar^2}{8}\{\!\{(h^{\uparrow\uparrow}-h^{\downarrow\downarrow}),f^{\uparrow\downarrow}\}\!\}
\nonumber\\
&-&h^{\uparrow\downarrow}(f^{\uparrow\uparrow}-f^{\downarrow\downarrow})
+\frac{i\hbar}{2}\{h^{\uparrow\downarrow},(f^{\uparrow\uparrow}+f^{\downarrow\downarrow})\}
+\frac{\hbar^2}{8}\{\!\{h^{\uparrow\downarrow},(f^{\uparrow\uparrow}-f^{\downarrow\downarrow})\}\!\}+....,
\nonumber\\
      i\hbar\dot f^{\downarrow\uparrow} &=&
f^{\downarrow\uparrow}(h^{\downarrow\downarrow}-h^{\uparrow\uparrow})
+\frac{i\hbar}{2}\{(h^{\downarrow\downarrow}+h^{\uparrow\uparrow}),f^{\downarrow\uparrow}\}
-\frac{\hbar^2}{8}\{\!\{(h^{\downarrow\downarrow}-h^{\uparrow\uparrow}),f^{\downarrow\uparrow}\}\!\}
\nonumber\\
&-&h^{\downarrow\uparrow}(f^{\downarrow\downarrow}-f^{\uparrow\uparrow})
+\frac{i\hbar}{2}\{h^{\downarrow\uparrow},(f^{\downarrow\downarrow}+f^{\uparrow\uparrow})\}
+\frac{\hbar^2}{8}\{\!\{h^{\downarrow\uparrow},(f^{\downarrow\downarrow}-f^{\uparrow\uparrow})\}\!\}+...,
\nonumber\\
      i\hbar\dot \kappa &=& \kappa\,(h^{\uparrow\uparrow}+\bar h^{\downarrow\downarrow})
  +\frac{i\hbar}{2}\{(h^{\uparrow\uparrow}-\bar h^{\downarrow\downarrow}),\kappa\}    
  -\frac{\hbar^2}{8}\{\!\{(h^{\uparrow\uparrow}+\bar h^{\downarrow\downarrow}),\kappa\}\!\}   
 \nonumber\\ 
 &+&\Delta\,(f^{\uparrow\uparrow}+\bar f^{\downarrow\downarrow})
  +\frac{i\hbar}{2}\{(f^{\uparrow\uparrow}-\bar f^{\downarrow\downarrow}),\Delta\}    
  -\frac{\hbar^2}{8}\{\!\{(f^{\uparrow\uparrow}+\bar f^{\downarrow\downarrow}),\Delta\}\!\}
  - \Delta + ...,
 \nonumber\\     
    i\hbar\dot \kappa^* &=&  -\kappa^*(h^{\uparrow\uparrow}+\bar h^{\downarrow\downarrow}) 
  +\frac{i\hbar}{2}\{(h^{\uparrow\uparrow}-\bar h^{\downarrow\downarrow}),\kappa^*\}    
  +\frac{\hbar^2}{8}\{\!\{(h^{\uparrow\uparrow}+\bar h^{\downarrow\downarrow}),\kappa^*\}\!\}   
  \nonumber\\
  &-& \Delta^*(f^{\uparrow\uparrow}+\bar f^{\downarrow\downarrow}) 
  +\frac{i\hbar}{2}\{(f^{\uparrow\uparrow}-\bar f^{\downarrow\downarrow}),\Delta^*\}    
  +\frac{\hbar^2}{8}\{\!\{(f^{\uparrow\uparrow}+\bar f^{\downarrow\downarrow}),\Delta^*\}\!\}
  + \Delta^* +..., 
\label{WHF}
\end{eqnarray} \end{widetext}
where the functions $h$, $f$, $\Delta$, and $\kappa$ are the Wigner
transforms of $\hat h$, $\hat\rho$, $\hat\Delta$, and $\hat\kappa$,
respectively, $\bar f(\br,\bp)=f(\br,-\bp)$,
 $\{f,g\}$ is the Poisson
bracket of the functions $f$ and $g$ and
$\{\{f,g\}\}$ is their double Poisson bracket.
The dots stand for terms proportional to higher powers of $\hbar$ -- after 
integration over phase space these terms disappear and we arrive to the set of
exact integral equations.
This set of equations must be complemented by the dynamical equations for 
$\bar f^{\uparrow\uparrow}, \bar f^{\downarrow\downarrow}, \bar f^{\uparrow\downarrow}, 
\bar f^{\downarrow\uparrow},\bar\kappa,\bar\kappa^*$.
They are obtained by the change $\bp \rightarrow -\bp$ in arguments of functions and Poisson brackets. 
So, in reality we deal with the set of twelve equations. We introduced the notation
$\kappa \equiv \kappa^{\uparrow\downarrow}$ and $\Delta \equiv \Delta^{\uparrow\downarrow}$.

Following the papers \cite{BaMo} in the next step we write above equations in terms of spin-scalar
\mbox{$f^+=f^{\uparrow\uparrow}+ f^{\downarrow\downarrow}$}
and spin-vector
\mbox{$f^-=f^{\uparrow\uparrow}- f^{\downarrow\downarrow}$}
functions. 
As a result, we obtain a set of twelve equations, which is solved by the method of moments in a small amplitude 
approximation. To this end 
all functions $f(\br,\bp,t)$ and $\kappa(\br,\bp,t)$ are divided into equilibrium part 
and deviation (variation): $f(\br,\bp,t)=f(\br,\bp)_{eq}+\delta f(\br,\bp,t)$, 
\mbox{$\kappa(\br,\bp,t)=\kappa(\br,\bp)_{eq}+\delta \kappa(\br,\bp,t)$}.
Then equations are linearized neglecting quadratic 
in $\delta f$ and $\delta \kappa$ terms~\cite{BaMoPRC}.

\subsection{Model Hamiltonian}

 The microscopic Hamiltonian of the model, harmonic oscillator with 
spin orbit potential plus separable quadrupole-quadrupole and 
spin-spin residual interactions is given by
\begin{eqnarray}
\label{Ham}
 H=\sum\limits_{i=1}^A\left[\frac{\hat\bp_i^2}{2m}+\frac{1}{2}m\omega^2\br_i^2
-\eta\hat \bl_i\hat \bS_i\right]+H_{qq}+H_{ss}\quad
\end{eqnarray}
with
\begin{eqnarray}
\label{Hqq}
H_{qq}&=&
\sum_{\mu=-2}^{2}(-1)^{\mu}
\left\{\bar{\kappa}
 \sum\limits_i^Z\!\sum\limits_j^N
+\frac{\kappa}{2}
\left[\sum\limits_{\substack{i,j \\ (i\neq j)}}^{Z}
+\sum\limits_{\substack{i,j \\ (i\neq j)}}^{N}
\right]
\right\}\quad
\nonumber \\
&\times& q_{2-\mu}(\br_i)q_{2\mu}(\br_j),
\nonumber\\
\label{Hss}
H_{ss}&=&
\sum_{\mu=-1}^{1}(-1)^{\mu}
\left\{\bar{\chi}
 \sum\limits_i^Z\!\sum\limits_j^N
+\frac{\chi}{2}
\left[
\sum\limits_{\substack{i,j \\ (i\neq j)}}^{Z}
+\sum\limits_{\substack{i,j \\ (i\neq j)}}^{N}
\right]
\right\}\quad
\nonumber \\
&\times& \hat S_{-\mu}(i)\hat S_{\mu}(j)
\,\delta(\br_i-\br_j),\nonumber
\end{eqnarray}
where $N$ and $Z$ are numbers of neutrons and protons,
$q_{2\mu}(\br)=\sqrt{16\pi/5}\,r^2Y_{2\mu}(\theta,\phi)$
and $\hat S_{\mu}$ are spin matrices \cite{Var}.

\subsection{Pair potential}
The Wigner transform of the pair potential (pairing gap) $\Delta(\br,\bp)$ is related to 
the Wigner transform of the anomalous density by \cite{Ring}
\begin{equation}
\Delta(\br,\bp)=-\int\! \frac{d^3p'}{(2\pi\hbar)^3}
v(|\bp-\bp'|)\kappa(\br,\bp'),
\label{DK}
\end{equation}
where $v(p)$ is a Fourier transform of the two-body interaction.
We take for the pairing interaction a simple Gaussian, 
$v(p)=\beta {\rm e}^{-\alpha p^2}$  \cite{Ring}
with $\beta=-|V_0|(r_p\sqrt{\pi})^3$ and $\alpha=r_p^2/4\hbar^2$. 
The following values of parameters were used in calculations:  
$r_p=1.9$~fm,
$|V_0|=25$~MeV.
Several exceptions were done for rare earth nuclei: $|V_0|=26$~MeV for $^{150}$Nd, $|V_0|=26.5$~MeV for $^{176, 178, 180}$Hf and $^{182, 184}$W, 
$|V_0|=27$~MeV for nuclei with deformation $\delta\leqslant 0.18$.

\subsection{Equations of motion}

Integrating the set of equations for the $\delta f_{\tau}^{\varsigma}(\br,\bp,t)$ 
and  $\delta \kappa_{\tau}(\br,\bp,t)$ over phase space with the weights 
\begin{equation}
W =\{r\otimes p\}_{\lambda\mu},\,\{r\otimes r\}_{\lambda\mu},\,
\{p\otimes p\}_{\lambda\mu}, \mbox{ and } 1
\label{weightfunctions}
\end{equation}
one gets dynamic equations for the following collective variables:
\begin{eqnarray}
&&\L^{\tau\varsigma}_{\lambda\mu}(t)=\int\! d(\bp,\br) \{r\otimes p\}_{\lambda\mu}
\delta f^{\varsigma}_\tau(\br,\bp,t),
\nonumber\\
&&\R^{\tau\varsigma}_{\lambda\mu}(t)=\int\! d(\bp,\br) \{r\otimes r\}_{\lambda\mu}
\delta f^{\varsigma}_\tau(\br,\bp,t),
\nonumber\\
&&\P^{\tau\varsigma}_{\lambda\mu}(t)=\int\! d(\bp,\br) \{p\otimes p\}_{\lambda\mu}
\delta f^{\varsigma}_\tau(\br,\bp,t),
\nonumber\\
&&\F^{\tau\varsigma}(t)=\int\! d(\bp,\br)
\delta f^{\varsigma}_\tau(\br,\bp,t),
\nonumber\\
&&\tilde{\L}^{\tau}_{\lambda\mu}(t)=\int\! d(\bp,\br) \{r\otimes p\}_{\lambda\mu}
\delta \kappa_\tau(\br,\bp,t),
\nonumber\\
&&\tilde{\R}^{\tau}_{\lambda\mu}(t)=\int\! d(\bp,\br) \{r\otimes r\}_{\lambda\mu}
\delta \kappa_\tau(\br,\bp,t),
\nonumber\\
&&\tilde{\P}^{\tau}_{\lambda\mu}(t)=\int\! d(\bp,\br) \{p\otimes p\}_{\lambda\mu}
\delta \kappa_\tau(\br,\bp,t),\qquad
\label{Varis}
\end{eqnarray}
where 
$\varsigma=+,\,-,\,\uparrow\downarrow,\,\downarrow\uparrow,\
\{r\otimes p\}_{\lambda\mu}=\sum\limits_{\sigma,\nu}C_{1\sigma,1\nu}^{\lambda\mu}r_{\sigma}p_{\nu}.$
It is convenient to rewrite the dynamical equations in terms
of isoscalar and isovector variables
\begin{eqnarray}
\label{Isovs}
&&\bar \R_{\lambda\mu}=\R_{\lambda\mu}^{n}+\R_{\lambda\mu}^{p},\quad
\R_{\lambda\mu}=\R_{\lambda\mu}^{n}-\R_{\lambda\mu}^{p},
\nonumber\\ 
&&\bar \P_{\lambda\mu}=\P_{\lambda\mu}^{n}+\P_{\lambda\mu}^{p},\quad
\P_{\lambda\mu}=\P_{\lambda\mu}^{n}-\P_{\lambda\mu}^{p},
\nonumber\\
&&\bar \L_{\lambda\mu}=\L_{\lambda\mu}^{n}+\L_{\lambda\mu}^{p},\quad
\L_{\lambda\mu}=\L_{\lambda\mu}^{n}-\L_{\lambda\mu}^{p}.\qquad
\end{eqnarray}
We are interested in the scissors mode with quantum number
$K^{\pi}=1^+$. Therefore, we only need the part of dynamic equations 
with $\mu=1$. The integration yields the following set of equations 
for isovector variables~\cite{BaMoPRC2}:
\begin{widetext}
\begin{eqnarray}
\label{iv}
     \dot {\L}^{+}_{21}&=&
\frac{1}{m}\P_{21}^{+}-
\left[m\,\omega^2
-4\sqrt3\alpha\kappa_0R_{00}^{\rm eq}
+\sqrt6(1+\alpha)\kappa_0 R_{20}^{\rm eq}\right]\R^{+}_{21}
-i\hbar\frac{\eta}{2}\left[\L_{21}^-
+2\L^{\uparrow\downarrow}_{22}+
\sqrt6\L^{\downarrow\uparrow}_{20}\right],
\nonumber\\
     \dot {\L}^{-}_{21}&=&
\frac{1}{m}\P_{21}^{-}
-
\left[m\,\omega^2+\sqrt6\kappa_0 R_{20}^{\rm eq}
-\frac{\sqrt{3}}{20}\hbar^2 
\left( \chi-\frac{\bar\chi}{3} \right)
\left(\frac{I_1}{a_0^2}+\frac{I_1}{a_1^2}\right)\left(\frac{a_1^2}{A_2}-\frac{a_0^2}{A_1}\right)
\right]\R^{-}_{21}
-i\hbar\frac{\eta}{2}\L_{21}^+ 
+\frac{4}{\hbar}|V_0| I_{rp}^{\kappa\Delta}(r') {\tilde\L}_{21},
\nonumber\\
     \dot {\L}^{\uparrow\downarrow}_{22}&=&
\frac{1}{m}\P_{22}^{\uparrow\downarrow}-
\left[m\,\omega^2-2\sqrt6\kappa_0R_{20}^{\rm eq}
-\frac{\sqrt{3}}{5}\hbar^2 
\left( \chi-\frac{\bar\chi}{3} \right)\frac{I_1}{A_2}
\right]\R^{\uparrow\downarrow}_{22}
-i\hbar\frac{\eta}{2}\L_{21}^+,
\nonumber\\
     \dot {\L}^{\downarrow\uparrow}_{20}&=&
\frac{1}{m}\P_{20}^{\downarrow\uparrow}-
\left[m\,\omega^2
+2\sqrt6\kappa_0 R_{20}^{\rm eq}\right]\R^{\downarrow\uparrow}_{20}
+4\sqrt3\kappa_0 R_{20}^{\rm eq}\,\R^{\downarrow\uparrow}_{00}
-i\hbar\frac{\eta}{2}\sqrt{\frac{3}{2}}\L_{21}^+ 
\nonumber\\
&&+\frac{\sqrt{3}}{15}\hbar^2 
\left( \chi-\frac{\bar\chi}{3} \right)I_1 \,
\left[
\left(\frac{1}{A_2}-\frac{2}{A_1}\right)
\R_{20}^{\downarrow\uparrow}+
\sqrt2
\left(\frac{1}{A_2}+\frac{1}{A_1}\right)
\R_{00}^{\downarrow\uparrow}
\right],
\nonumber\\
     \dot {\L}^{+}_{11}&=&
-3\sqrt6(1-\alpha)\kappa_0 R_{20}^{\rm eq}\,\R^{+}_{21}
-i\hbar\frac{\eta}{2}\left[\L_{11}^- 
+\sqrt2\L^{\downarrow\uparrow}_{10}\right],
\nonumber\\
     \dot {\L}^{-}_{11}&=&
-\left[3\sqrt6\kappa_0 R_{20}^{\rm eq}
-\frac{\sqrt{3}}{20}\hbar^2 
\left( \chi-\frac{\bar\chi}{3} \right)
\left(\frac{I_1}{a_0^2}-\frac{I_1}{a_1^2}\right)\left(\frac{a_1^2}{A_2}-\frac{a_0^2}{A_1}\right)
\right]\R^{-}_{21}
-\hbar\frac{\eta}{2}\left[i\L_{11}^+
+\hbar \F^{\downarrow\uparrow}\right]
+\frac{4}{\hbar}|V_0| I_{rp}^{\kappa\Delta}(r') {\tilde\L}_{11},
\nonumber\\
     \dot {\L}^{\downarrow\uparrow}_{10}&=&
-\hbar\frac{\eta}{2\sqrt2}\left[i\L_{11}^+
+\hbar \F^{\downarrow\uparrow}\right],
\qquad\qquad
     \dot { \F}^{\downarrow\uparrow} = 
-\eta\left[\L_{11}^- +\sqrt2\L^{\downarrow\uparrow}_{10}\right],
\nonumber\\
\dot {\R}^{+}_{21}&=&
\frac{2}{m}\L_{21}^{+}
-i\hbar\frac{\eta}{2}\left[\R_{21}^-
+2\R^{\uparrow\downarrow}_{22}+
\sqrt6\R^{\downarrow\uparrow}_{20}\right],
\nonumber\\
     \dot {\R}^{-}_{21}&=&
\frac{2}{m}\L_{21}^{-}
-i\hbar\frac{\eta}{2}\R_{21}^+,
\nonumber\\
     \dot {\R}^{\uparrow\downarrow}_{22}&=&
\frac{2}{m}\L_{22}^{\uparrow\downarrow}
-i\hbar\frac{\eta}{2}\R_{21}^+,
\qquad\qquad
     \dot {\R}^{\downarrow\uparrow}_{20} = 
\frac{2}{m}\L_{20}^{\downarrow\uparrow}
-i\hbar\frac{\eta}{2}\sqrt{\frac{3}{2}}\R_{21}^+,
\nonumber\\
     \dot {\P}^{+}_{21}&=&
-2\left[m\,\omega^2+\sqrt6\kappa_0 R_{20}^{\rm eq}\right]\L^{+}_{21}
+6\sqrt6\kappa_0 R_{20}^{\rm eq}\L^{+}_{11}
-i\hbar\frac{\eta}{2}\left[\P_{21}^- 
+2\P^{\uparrow\downarrow}_{22}+\sqrt6\P^{\downarrow\uparrow}_{20}\right]
\nonumber\\
&&+\frac{3\sqrt{3}}{4}\hbar^2 
\chi \frac{I_2}{A_1A_2}
\left[\left(A_1-A_2\right) \L_{21}^{+} +
\left(A_1+A_2\right) \L_{11}^{+}\right]
+\frac{4}{\hbar}|V_0| I_{pp}^{\kappa\Delta}(r') {\tilde\P}_{21},
\nonumber\\
     \dot {\P}^{-}_{21}&=&
-2\left[m\,\omega^2+\sqrt6\kappa_0 R_{20}^{\rm eq}\right]\L^{-}_{21}
+6\sqrt6\kappa_0 R_{20}^{\rm eq}\L^{-}_{11}
-6\sqrt2\kappa_0 L_{10}^-(\rm eq)\R^{+}_{21}
-i\hbar\frac{\eta}{2}\P_{21}^{+}
\nonumber\\
&&+\frac{3\sqrt{3}}{4}\hbar^2 
\chi \frac{I_2}{A_1A_2}
\left[\left(A_1-A_2\right)\L_{21}^{-} +
\left(A_1+A_2\right) \L_{11}^{-}\right],
\nonumber\\
     \dot {\P}^{\uparrow\downarrow}_{22}&=&
-\left[2m\,\omega^2-4\sqrt6\kappa_0 R_{20}^{\rm eq}
-\frac{3\sqrt{3}}{2}\hbar^2 
\chi \frac{I_2}{A_2}
\right]\L^{\uparrow\downarrow}_{22}
-i\hbar\frac{\eta}{2}\P_{21}^{+},
\nonumber\\
     \dot {\P}^{\downarrow\uparrow}_{20}&=&
-\left[2m\,\omega^2+4\sqrt6\kappa_0 R_{20}^{\rm eq}\right]\L^{\downarrow\uparrow}_{20}
+8\sqrt3\kappa_0 R_{20}^{\rm eq}\L^{\downarrow\uparrow}_{00}
-i\hbar\frac{\eta}{2}\sqrt{\frac{3}{2}}\P_{21}^{+}
\nonumber\\
&&+\frac{\sqrt{3}}{2}\hbar^2 
\chi \frac{I_2}{A_1A_2}
\left[\left(A_1-2A_2\right)\L_{20}^{\downarrow\uparrow}+
\sqrt2\left(A_1+A_2\right) \L_{00}^{\downarrow\uparrow}
\right],
\nonumber\\
     \dot {\L}^{\downarrow\uparrow}_{00}&=&
\frac{1}{m}\P_{00}^{\downarrow\uparrow}-m\,\omega^2\R^{\downarrow\uparrow}_{00}
+4\sqrt3\kappa_0 R_{20}^{\rm eq}\,\R^{\downarrow\uparrow}_{20}
\nonumber\\
&&+\frac{1}{2\sqrt{3}}\hbar^2 
\left[\left( \chi-\frac{\bar\chi}{3} \right)I_1-\frac{9}{4}\chi I_2\right]
\left[\left(\frac2{A_2}-\frac1{A_1}\right)\R_{00}^{\downarrow\uparrow}+
\sqrt2\left(\frac1{A_2}+\frac1{A_1}\right)\R_{20}^{\downarrow\uparrow}
\right],
\nonumber\\
     \dot {\R}^{\downarrow\uparrow}_{00}&=&
\frac{2}{m}\L_{00}^{\downarrow\uparrow},
\nonumber
\\
     \dot {\P}^{\downarrow\uparrow}_{00}&=&
-2m\,\omega^2\L^{\downarrow\uparrow}_{00}
+8\sqrt3\kappa_0 R_{20}^{\rm eq}\,\L^{\downarrow\uparrow}_{20}
+\frac{\sqrt{3}}{2}\hbar^2 
\chi I_2
\left[\left(\frac2{A_2}-\frac1{A_1}\right)\L_{00}^{\downarrow\uparrow}+
\sqrt2\left(\frac1{A_2}+\frac1{A_1}\right)\L_{20}^{\downarrow\uparrow}
\right],
 \nonumber\\
     \dot{ {\tilde\P}}_{21} &=& -\frac{1}{\hbar}\Delta_0(r') \P^+_{21} + 6\hbar\alpha\kappa_0 K_0{\cal R}^+_{21}, 
\nonumber\\
     \dot {{\tilde\L}}_{21} &=& -\frac{1}{\hbar}\Delta_0(r') {\L}^-_{21},
\qquad\qquad
     \dot {{\tilde\L}}_{11} = -\frac{1}{\hbar}\Delta_0(r') \L^-_{11},
\end{eqnarray}
where 
\begin{eqnarray}
&&A_1=\sqrt2\, R_{20}^{\rm eq}-R_{00}^{\rm eq}=\frac{Q_{00}}{\sqrt3}\left(1+\frac{4}{3}\delta\right),
\quad A_2= R_{20}^{\rm eq}/\sqrt2+R_{00}^{\rm eq}=-\frac{Q_{00}}{\sqrt3}\left(1-\frac{2}{3}\delta\right),
\nonumber
\end{eqnarray}
\begin{eqnarray}
&&a_{-1} = a_1 = R_0\left( \frac{1-(2/3)\delta}{1+(4/3)\delta} \right)^{1/6},
\quad a_0 = R_0\left( \frac{1-(2/3)\delta}{1+(4/3)\delta} \right)^{-1/3},
\quad Q_{00}=\frac{3}{5}AR_0^2
\nonumber
\end{eqnarray}
$a_i$ are semiaxes of ellipsoid by which the shape of nucleus is approximated, $\delta$ -- deformation parameter, 
$R_0=1.2A^{1/3}$~fm -- radius of nucleus.
$$I_1=\frac{\pi}{4}\int\limits_{0}^{\infty}dr\, r^4\left(\frac{\partial n(r)}{\partial r}\right)^2,
\quad
I_2=\frac{\pi}{4}\int\limits_{0}^{\infty}dr\, r^2 n(r)^2,$$
$n(r)$ -- nuclear density.
The values $L_{10}^-(\rm eq)$, $K_0$, $\Delta_0(r')$, $I_{rp}^{\kappa\Delta}(r')$, 
$I_{pp}^{\kappa\Delta}(r')$ entering into the equations~(\ref{iv}), details of calculations relating to accounting for pair correlations
and choice of parameters
are discussed in the Ref.~\cite{BaMoPRC2}.

\subsection{Energies and excitation probabilities}
   
Imposing the time evolution via $\di{e^{i\Omega t}}$ for all variables
one transforms (\ref{iv}) into a set of algebraic equations. 
Eigenfrequencies are found as the zeros of its secular equation. Excitation
probabilities are calculated with the help of the theory of linear 
response of the system to a weak external field
\begin{equation}
\label{extf}
\hat O(t)=\hat O\,\e^{-i\Omega t}+\hat O^{\dagger}\,e^{i\Omega t}.
\end{equation}
A detailed explanation can be found in \cite{BaMo,BS}. 
We recall only the main points.
The matrix elements of the operator $\hat O$ obey the relationship \cite{Lane}
\begin{equation}
\label{matel}
|\langle\psi_a|\hat O|\psi_0\rangle|^2=
\hbar\lim_{\Omega\to\Omega_a}(\Omega-\Omega_a)
\overline{\langle\psi'|\hat O|\psi'\rangle\e^{-i\Omega t}},
\end{equation}
where $\psi_0$ and $\psi_a$ are the stationary wave functions of the
unperturbed ground and excited states; $\psi'$ is the wave function
of the perturbed ground state, $\Omega_a=(E_a-E_0)/\hbar$ are the
normal frequencies, the bar means averaging over a time interval much
larger than $1/\Omega$.

To calculate the magnetic transition probability, it is necessary
to excite the system by the following external field:
\begin{equation}
\label{Magnet}
\hat O_{\lambda\mu}=\mu_N\left(g_s\hat\bS/\hbar-ig_l\frac{2}{\lambda+1}[\br\times\nabla]\right)
\nabla(r^{\lambda}Y_{\lambda\mu}), \quad 
\mu_N=\frac{e\hbar}{2mc}.
\end{equation}
Here $g_l^{\rm p}=1,$ $g_s^{\rm p}=5.5856$ for protons and $g_l^{\rm n}=0,$ $g_s^{\rm n}=-3.8263$ for neutrons.
The dipole operator \mbox{($\lambda=1,\ \mu=1$)} 
in cyclic coordinates looks like
\begin{equation}
\label{Magnet_1}
\hat O_{11}=
\mu_N\sqrt{\frac{3}{4\pi}}\left[g_s\hat S_{1}/\hbar
-g_l\sqrt2\sum_{\nu,\sigma}
C_{1\nu,1\sigma}^{11}r_{\nu}\nabla_{\sigma}\right].
\end{equation}
Its Wigner transform is
\begin{equation}
\label{MagWig}
(\hat O_{11})_W=
\sqrt{\frac{3}{4\pi}}\left[g_s\hat S_{1}
-ig_l\sqrt2\sum_{\nu,\sigma}
C_{1\nu,1\sigma}^{11}r_{\nu}p_{\sigma}\right]\frac{\mu_N}{\hbar}.
\end{equation}
 For the matrix element we have
\begin{eqnarray}
\label{psiO}
\langle\psi'|\hat O_{11}|\psi'\rangle &=&
\sqrt{\frac{3}{2\pi}}\left[-\frac{\hbar}{2}
(g_s^{\rm n}\F^{\rm n\d}+g_s^{\rm p}\F^{\rm p\d})
-ig_l^{\rm p}\L_{11}^{\rm p+}\right]\frac{\mu_N}{\hbar}
\nonumber\\
&=&\sqrt{\frac{3}{8\pi}}\left[-\frac{1}{2}
[(g_s^{\rm n}-g_s^{\rm p})\F^{\d}+(g_s^{\rm n}+g_s^{\rm p})\bar \F^{\d}]
-\frac{i}{\hbar}g_l^{\rm p}(\bar\L_{11}^{+}- \L_{11}^{+})\right]\mu_N
\nonumber\\
&=&\sqrt{\frac{3}{8\pi}}\left[
\frac{1}{2}(g_s^{\rm p}-g_s^{\rm n})\F^{\downarrow\uparrow}
+\frac{i}{\hbar}g_l^{\rm p} \L_{11}^{+}
+\frac{i}{\hbar}[g_s^{\rm n}+g_s^{\rm p}-g_l^{\rm p}]\bar\L_{11}^{+}
\right]\mu_N.
\end{eqnarray}
 Deriving (\ref{psiO}) we have used the relation $2i\bar\L^+_{11}=-\hbar \bar\F^{\d}$,
which follows from the angular momentum conservation~\cite{BaMo}.

One has to add the external field (\ref{Magnet_1}) to the Hamiltonian (\ref{Ham}). 
Due to the external field some dynamical equations of 
(\ref{iv}) become inhomogeneous:
\begin{eqnarray}
     \dot \R^{+}_{21}
&=&\ldots\; +i\frac{3}{\sqrt{\pi}}\frac{\mu_N}{4\hbar} g^{\rm p}_l R^{+}_{20}(\rm eq)\,\e^{i\Omega t},
\nonumber\\
     \dot \L^{-}_{11}
&=&\ldots\; + i\sqrt{\frac{3}{\pi}}\frac{\mu_N}{4\hbar} g^{\rm p}_l L^{-}_{10}(\rm eq)\,\e^{i\Omega t},
\nonumber\\
     \dot \L^{\d}_{10}
&=&\ldots\; 
+i\sqrt{\frac{3}{2\pi}}\frac{\mu_N}{4\hbar}\left( g^{\rm n}_s-g^{\rm p}_s\right) L^{-}_{10}(\rm eq)  \e^{i\Omega t}.
\end{eqnarray}
For the isoscalar set of equations, respectively, we obtain:
\begin{eqnarray}
     \dot {\bar\R}^{+}_{21}
&=&\ldots\; - i\frac{3}{\sqrt{\pi}}\frac{\mu_N}{4\hbar} g^{\rm p}_l R^{+}_{20}(\rm eq)\,\e^{i\Omega t},
\nonumber\\
     \dot {\bar\L}^{-}_{11}
&=&\ldots\; - i\sqrt{\frac{3}{\pi}}\frac{\mu_N}{4\hbar} g^{\rm p}_l L^{-}_{10}(\rm eq)\,\e^{i\Omega t},
\nonumber\\
     \dot {\bar\L}^{\d}_{10}
&=&\ldots\; 
+i\sqrt{\frac{3}{2\pi}}\frac{\mu_N}{4\hbar}\left( g^{\rm n}_s+g^{\rm p}_s\right) L^{-}_{10}(\rm eq) \e^{i\Omega t}.
\end{eqnarray}\end{widetext}
Solving the inhomogeneous set of equations 
one can find the required in (\ref{psiO}) values of
$\L_{11}^{+}$ , $\bar\L_{11}^{+}$ and $\F^{\d}$ and using (\ref{matel}) calculate 
$B(M1)$ factors for all excitations.

One also should be aware of the fact that straightforward application of 
Lane's formula to the present WFM approach which leads to non-symmetric 
eigenvalue problems may yield negative transition 
probabilities violating the starting relation~(\ref{matel}). However, with the 
parameters employed here and also in our previous works~\cite{BaMo,BaMoPRC,BaMoPRC2,Malov,Urban,BS}, 
this never happened.

\end{document}